\documentclass[a4paper,11pt]{article}
\pdfoutput=1 
\usepackage{soul}

\usepackage{jcappub} 

\usepackage[T1]{fontenc} 
\usepackage[normalem]{ulem}

\newcommand{\mr}[1]{\mathrm{#1}}
\newcommand{\mf}[1]{\mathbf{#1}}
\newcommand{\tf}[1]{\textbf{#1}}
\newcommand{\ol}[1]{\overline{#1}}
\newcommand{\wh}[1]{\widehat{#1}}

\usepackage{cancel}
\usepackage{multirow}
\usepackage{mathtools}

\title{\boldmath 
Joint analysis of anisotropic power spectrum, bispectrum and trispectrum: application to N-body simulations}

\author[a,b]{Davide Gualdi,}
\author[a,b]{H\'ector Gil-Mar\'in}
\author[a,c]{and Licia Verde}

\affiliation[a]{Institut de Ci\`encies del Cosmos, University of Barcelona, ICCUB, Barcelona 08028, Spain}
\affiliation[b]{Institute of Space Studies of Catalonia (IEEC), E-08034 Barcelona, Spain}
\affiliation[c]{Instituci\'o Catalana de Recerca i Estudis Avan\c{c}ats, Passeig Llu\'is Companys 23, Barcelona 08010, Spain}

\emailAdd{dgualdi@icc.ub.edu}
\emailAdd{hectorgil@icc.ub.edu}
\emailAdd{liciaverde@icc.ub.edu}

\abstract{ We perform for the first time a joint analysis of the monopole and quadrupoles for power spectrum, bispectrum and integrated trispectrum (i-trispectrum) from the redshift space matter field in N-body simulations. With a full Markov Chain Monte Carlo exploration of the posterior distribution, we quantify the constraints on cosmological parameters for an object density of $n_\mr{p}=5\times10^{-4} (h\,\mr{Mpc}^{-1})^{3}$, redshift $z=0.5$, and a covariance corresponding to a survey volume of $V_\mr{survey} =25\,(h^{-1}\mr{Gpc})^3$, a set up which is representative of forthcoming galaxy redshift surveys. We demonstrate the complementarity of the bispectrum and i-trispectrum in constraining key cosmological parameters. In particular, compared to the state-of-the-art power spectrum (monopole plus quadrupole) and bispectrum (monopole) analyses, we find 1D $68\%$ credible regions smaller by a factor of $(72\%,78\%,72\%,47\%,46\%)$ for the parameters $(f,\sigma_8,f_\mr{nl},\alpha_\parallel,\alpha_\perp)$ respectively. This work motivates the additional effort necessary to include the redshift-space anisotropic signal of higher-order statistics in the analysis and interpretation of ongoing and future galaxy surveys. }

\begin{document}
\maketitle
\flushbottom

\section{Introduction}
The matter distribution statistical properties at large scales provide a unique window to both constrain key cosmological parameters and test cosmological models.
A key approach to study these large scale structures (LSS) properties is to trace the matter distribution by measuring galaxy redshifts and angular positions in the sky. This is the target of current and future LSS surveys (e.g., DESI\footnote{\url{http://desi.lbl.gov}} \citep{Levi:2013gra}; Euclid \footnote{\url{http://sci.esa.int/euclid/}} \citep{Laureijs:2011gra}; PFS \footnote{\url{http://pfs.ipmu.jp}} \citep{Ellis:2012rn}; SKA\footnote{\url{https://www.skatelescope.org}} \citep{Bacon:2018dui}; LSST\footnote{\url{https://www.lsst.org/}} \citep{Abell:2009aa} and WFIRST\footnote{\url{https://www.cosmos.esa.int/web/wfirst}} \cite{Green:2012mj}).

In the last thirty years the two-point correlation function in configuration space \cite{Peebles1980} and its Fourier counterpart, the power spectrum, have been the main statistical tools used to interpret LSS data. Both kind of analyses have produced unprecedented constraints on cosmological models using only late-time (low redshifts) observations. Nevertheless, two-point statistics are not optimally suited to capture mode-coupling or non-Gaussianities: for this, higher-order statistics are usually employed.

The study and analysis of three-point statistics has long been recognised as a promising avenue
\cite{Groth:1977gj,1975ApJ...196....1P,1982ApJ...259..474F,Fry:1983cj,Matarrese:1997sk,Verde:1998zr,Verde:2001sf,Scoccimarro:1997st,Scoccimarro:2000ee,Scoccimarro:1999ed,Scoccimarro:2000sp} motivating further developments in recent years including improving the theoretical modelling by adding loop corrections \cite{Sefusatti:2009qh,Sefusatti:2011gt,Hashimoto:2017klo,Desjacques:2018pfv,Eggemeier:2018qae,Castiblanco:2018qsd,Eggemeier:2021cam}.
 Originally, the bispectrum --the three-point correlation function Fourier space counterpart-- was introduced to break the degeneracies between bias and cosmological parameters \cite{Fry:1992vr,Fry:1993bj}.
Recently however, it has been shown to generally add significant constraining power when combined with the power spectrum \cite{Yankelevich:2018uaz,Oddo:2019run,Barreira:2019icq,Agarwal:2020lov}.
In particular the bispectrum shows promising potential in constraining neutrinos masses \cite{Ruggeri:2017dda,Coulton:2018ebd,Hahn:2019zob,Hahn:2020lou,Kamalinejad:2020izi}, detecting specific signatures of general relativity and relativistic effects \cite{GilMarin:2011xq,Bartolo:2013ws,Bellini:2015wfa,Bertacca:2017dzm,DiDio:2018unb}, and measuring baryonic acoustic oscillations (BAO) \cite{Pearson:2017wtw,Child:2018klv}, as it is also the case for the three-point correlation function \cite{Slepian:2015hca,Slepian:2016kfz}.
Moreover, when considering primordial non-Gaussianities, three-point statistics are essential in lifting degeneracies with other cosmological parameters \citep{Verde:1999ij,Scoccimarro:2003wn,Jeong:2009vd,Bose:2018zpk,Karagiannis:2018jdt}.

Also in weak-lensing analyses the three-point correlation function and the bispectrum have been widely studied in the literature since the first seminal works \citep{Takada:2003ef,Kilbinger:2005jy,Semboloni:2010er}. More recent works focused on determining their added constraining potential with respect to two-point statistics \cite{Kayo:2012nm,Rizzato:2018whp,Coulton:2018ebd,Halder:2021itp,Jung:2021weh,Pyne:2020ijd}, together with application on data \cite{Fu:2014loa}.

Currently only the isotropic bispectrum signal has been used to constrain cosmological parameters from galaxy surveys data (see e.g, \cite{Gil-Marin:2014sta,Gil-Marin:2016wya} for the SDSSIII BOSS survey). Modelling of the anisotropic component has been proposed in several works \cite{Scoccimarro:2015bla,Sugiyama:2018yzo}, including studying its impact in terms of additional signal \cite{Sugiyama:2019ike} and tighter parameter constraints \cite{Gagrani:2016rfy,Gualdi:2020ymf}.

At the next higher-order level, the trispectrum --the four-point correlation function connected part in Fourier space-- has been successfully modelled in detail and measured in the context of the cosmic microwave background (CMB) fluctuations \cite{kunzetal2001,Komatsu:2002db,deTroia2003,Munshi:2009wy,Kamionkowski:2010me,Izumi:2011di,Regan_2015, Feng:2015pva,Fergusson:2010gn,Smith:2015uia,Namikawa:2017uke,PlanckTrispectrum18, Akrami:2019izv}.
In LSS studies it has received less attention \cite{Verde:2001pf,Cooray:2008eb,Lazeyras:2017hxw,Bellomo:2018lew}, because of the difficulties implied by both modelling and measuring the signal in three dimensions.
That is the reason why trispectrum applications to real and simulated data are scarce in the literature \cite{Fry1978,Suto:1993ua,Sabiu:2019kbh}.

Of particular interest is the trispectrum describing the non-Gaussian mode coupling inducing contributions to the power spectrum covariance matrix
\cite{2010A&A...514A..79P,2012JCAP...04..019D, Mohammed:2016sre,Taruya:2020qoy}.
In terms of modelling, the effective field theory formalism was applied to the trispectrum by \cite{Bertolini:2016bmt} and calibrated recently at 1-loop by \cite{Steele:2021lnz}; an angular coordinates formalism has been recently introduced by \cite{Lee:2020ebj}.
In the presence of a primordial trispectrum, the correction to the non-Gaussian linear bias was derived in \cite{Lazeyras:2015giz}. 

CMB analyses have proved the trispectrum constraining power for primordial non-Gaussianities, with implications in confirming or ruling out single/multi field inflation models \cite{Bartolo:2004if}. Since the late-time 3D matter field trispectrum contains by definition
more modes than the primordial 2D CMB counterpart, as pointed out by \cite{Verde:2001pf}, the LSS trispectrum can be a powerful tool in deriving late-time constraints on primordial non-Gaussianity.

An estimator for an integrated version of the 3D LSS trispectrum was proposed first by \cite{Sefusatti:2004xz} and in \cite{Gualdi:2020eag} we measured and modelled its isotropic signal in both real and redshift space. By measuring the "i-trispectrum" and its covariance matrix from the Quijote simulations suite \cite{Villaescusa-Navarro:2019bje} and through a Fisher forecast approach, we proved its potential in improving primordial non-Gaussianity constraints.

In this work for the first time the anisotropic redshift space power spectrum, bispectrum and i-trispectrum statistics are considered together in a single, joint data-vector. After modelling and measuring all quantities from simulations up to the quadrupole level, we 
quantify the additional constraining power of each term and their complementarity with a full Markov Chain Monte Carlo exploration of the posterior of the relevant cosmological parameters.

\renewcommand{\arraystretch}{2.}
\begin{table}[tbp]
\centering
\begin{tabular}{c|c|c|c|c}
\cline{1-5}
survey & volume & 
tracers & expected S/N $\bar{n}_\mr{g}P_\mr{g}(k)$ & redshift interval \\
\cline{1-5}
BOSS + eBOSS
&  12 $\mathrm{Gpc}^3$ & 2,248,436 & - & $0.07<z<2.2$ \\
DESI 
&  $\sim$65 $\mathrm{Gpc/h}^3$  & - & $\left[0.05-146.6\right]$ $(0.2\,h/\mr{Mpc})$ & $0.05<z<1.85$\\
PFS
& 6.6 $\mathrm{Gpc/h}^3$ & - & $\left[1-1.4\right]$ $(0.1\,h/\mr{Mpc})$ & $0.6<z<1.6$\\
\hline
\end{tabular}
\caption{\label{tab:surveys_specs} Volumes, number of tracers, expected signal to noise ratio and observed redshift intervals for current and future LSS clustering surveys: BOSS + eBOSS, DESI \cite{Aghamousa:2016zmz} and PFS. 
}
\end{table}

For this analysis to be relevant for
current and future galaxy clustering data-sets in terms of statistical error and signal-to-noise, we randomly sub-sample the matter particles in the simulations down to an object density of $n_\mr{p}=5\times 10^{-4}$ $(h\mr{Mpc}^{-1})^3$, together with re-scaling the covariance matrix to simulate an effective volume of $25$ $(h^{-1}\mr{Gpc})^3$. As a reference in Table \ref{tab:surveys_specs} we report observed volume, number of tracers, expected signal to noise and redshift range for BOSS + eBOSS\footnote{\url{http://sdss.org/science/final-bao-and-rsd-measurements/}}, DESI \cite{Aghamousa:2016zmz} and PFS\footnote{\url{http://member.ipmu.jp/masahiro.takada/pfs_whitepaper.pdf}} surveys.

The rest of the paper is organised as follows. The methodology is presented in Section \ref{sec:methodology}, where the simulations (\ref{sec:simulations}), 
the theoretical models for the statistics (\ref{sec:theo_models}), the chosen model's parameters (\ref{sec:choice_params}) and the multipole expansion (\ref{sec:multipoles_exp}) are described.
We conclude this section with a brief recap of the estimator used to measure all the data-vector terms (\ref{sec:estim}), a formalism review for both signal-to-noise ratio and Fisher forecasts analysis (\ref{sec:sn_an_fisher}), and the analysis description (\ref{sec:analysis_setup}).

In Section \ref{sec:results} the results are reported starting with testing the fit quality for each term's model against the average measurements from the simulations (\ref{sec:model_fit}).
A preliminary study of the additional constraining power given by each individual statistic is performed in terms of signal-to-noise ratio and Fisher forecasts in Section \ref{sec:snratio_res}.

The forecasts accuracy is compared to a standard MCMC sampling of the parameters posterior distribution for the possible data-vector's combinations in Section \ref{sec:mcmc_sampling}.
We conclude in Section \ref{sec:conclusions}.

\section{Methodology}
\label{sec:methodology}
After describing the N-body simulations used throughout the paper,
the adopted theoretical model for the statistics we use is presented here. The modelling of the power spectrum and bispectrum multipoles is taken from the literature and reported here for completeness. The (real space) i-trispectrum was presented in \cite{Gualdi:2020eag} but the i-trispectrum multipoles are presented here for the first time. 

For preliminary assessment of the constraining power of the statistics considered, we resort to the signal-to-noise ratio and to Fisher-based forecasts; these well-known methodologies are also reported in this section for completeness. In the results section we resort to Monte Carlo Markov Chains (MCMCs) and adopt the publicly available code \cite{2013PASP..125..306F}; this is a widely used tool and we refer the readers to the reference for further information. 

\subsection{Simulations}
\label{sec:simulations}
We measure the statistics of interest from the 8000 realisations of the \textsc{Quijote} N-body simulations suite \cite{Villaescusa-Navarro:2019bje}. 
Such a large number of realisations is necessary to accurately estimate the covariance matrix for the full data-vector including power spectrum, bispectrum and i-trispectrum multipoles.In this work the largest data-vector considered has 3615 elements.

In each simulation cubic box of side $L=1\, h^{-1}\textrm{Gpc}$, $512^3$ dark matter particles evolve from the initial conditions at $z=127$ (set using 2LPT \cite{Springel:2002uv,Crocce:2006ve,Scoccimarro:2011pz}) down to $z=0$ through gravitational collapse. Our analysis is focused on the snapshot at redshift $z=0.5$.
In terms of clustering properties, \cite{Villaescusa-Navarro:2019bje} shows an agreement between the power spectra measured from simulations with $512^3$ (fiducial) and $1024^3$ (high-resolution) with differences below the $2\%$ level up to $k=0.7$ $h/\mr{Mpc}$ at $z=0.5$. This accuracy shows that the adopted size of the box is well suited to our application which stops at much larger scales. 

The simulations cosmology of the \textsc{Quijote} suite is a flat $\Lambda$CDM model (consistent with the latest CMB constraints \cite{collaboration2018planck}). Specifically, the matter and baryon density parameters are $\Omega_\mr{m}=0.3175$, $\Omega_\mr{b}=0.049$, and the dark energy equation of state parameter is $w=-1$; the reduced Hubble parameter is $h\equiv H_0 / (100 \,\mr{km}\, \mr{s}^{-1}\mr{Mpc}^{-1})=0.6711$ ($H_0$ is the Hubble constant), the late-time dark matter fluctuations amplitude parameter is $\sigma_8=0.834$, the scalar spectral index is $n_\mr{s}=0.9624$, and neutrinos are massless, i.e. $M_\nu=0.0$ eV.

\subsection{Theoretical modelling}
\label{sec:theo_models}
The galaxy power spectrum model is computed using standard perturbation theory at two loops (2LPT) together with the TNS correcting factors \cite{Taruya:2010mx,Nishimichi:2011jm}, using the same code from \cite{Gil-Marin:2014sta}:

\begin{eqnarray}
P_\mr{g}(k,\mu) =
D^P_\mr{FoG}(k,\mu,\sigma_P)\times
\Big\{&&\!\!\!\!\!\!\!\!
P_{\mr{g},\delta\delta}(k) + 2f\mu^2P_{\mr{g}\delta\theta}(k)
+ f^2\mu^4 P_{\mr{g},\theta\theta}(k)
\notag \\
&+& b_1^3A^\mr{TNS}(k,\mu,f/b_1) 
+ b_1^4B^\mr{TNS}(k,\mu,f/b_1)
\Big\},
\end{eqnarray}

\noindent where $D^P_\mr{FoG}(k,\mu,\sigma_P)$ is the Lorentzian damping function, parametrised by $\sigma_P$, used to model the "Fingers of God" (FoG) effect due to redshift space distortions; $P_{\mr{g},\delta\delta}$, $P_{\mr{g},\delta\theta}$ and $P_{\theta\theta}$ are the galaxy density, galaxy density - velocity divergence and velocity divergence power spectra, respectively;
$A^\mr{TNS}$ and $B^\mr{TNS}$ encode the coupling between Kaiser and FoG effects \cite{Taruya:2010mx}. Finally, $\mu$ is the cosine of the angle between the $k$-vector $\mf{k}$ and the line of sight, while $b_1$ is the linear bias parameter and $f$ is the linear growth rate.

For both galaxy bispectrum and trispectrum we use the non-linear matter density power spectrum $P_{\delta\delta}$ as input, which is given at 2-loops order by \cite{Bernardeau:2001qr,Crocce:2005xy}

\begin{eqnarray}
P_{\delta\delta} =P^\mr{lin}
+2P^{(13)}_{\delta\delta}
+P^{(22)}_{\delta\delta}
+2P^{(15)}_{\delta\delta}
+2P^{(24)}_{\delta\delta}
+P^{(33)}_{\delta\delta}\,.
\end{eqnarray}

\noindent $P^\mr{lin}$ is the linear matter power spectrum, for which we use the CLASS code output \cite{Lesgourgues:2011re}. For each term $P^{(ij)}_{\delta\delta}$, $i$ and $j$ correspond to the perturbation expansion's order for each of the two $\delta$'s. 

Inspired by the findings of \cite{GilMarin:2011ik} which improve the performance and extend the regime of validity of the second-order perturbation theory (SPT) model for the bispectrum 
without increasing the computational time, we fit, for bispectrum and i-trispectrum separately, the coefficients of the tree-level-inspired kernels for the density and velocity divergence fields to the \textsc{Quijote} simulations, as explained in Appendix \ref{sec:app_fit_ker}. In particular, we fit the parameters\footnote{This is a simplified functional form compared to \cite{GilMarin:2011ik}, but, as we will show, it is sufficient for our purpose. Differently from \cite{GilMarin:2011ik} we did not check the redshift dependence of the fitted parameters $(f_1,f_2,f_3,g_1,g_2,g_3)$ since it is beyond the purpose of this work.} $(f_1,f_2,f_3,g_1,g_2,g_3)$, hence the acronym FPT: 

\begin{eqnarray}
F^{(2)}_\mr{FPT}\left[\mf{k}_a,\mf{k}_b\right]&=&
f_1\,\dfrac{5}{7} + 
f_2\,\dfrac{1}{2}\dfrac{\mf{k}_a\cdot\mf{k}_b}{k_ak_b}\left(\dfrac{k_a}{k_b}+\dfrac{k_b}{k_a}\right)+
f_3\,\dfrac{2}{7}\dfrac{(\mf{k}_a\cdot\mf{k}_b)^2}{k_a^2k_b^2}
\notag \\
G^{(2)}_\mr{FPT}\left[\mf{k}_a,\mf{k}_b\right]&=&
g_1\,\dfrac{3}{7} + 
g_2\,\dfrac{1}{2}\dfrac{\mf{k}_a\cdot\mf{k}_b}{k_ak_b}\left(\dfrac{k_a}{k_b}+\dfrac{k_b}{k_a}\right)+
g_3\,\dfrac{4}{7}\dfrac{(\mf{k}_a\cdot\mf{k}_b)^2}{k_a^2k_b^2}\,.
\label{eq:2SPT-likekernelsfit}
\end{eqnarray}

\noindent The standard perturbation theory kernels derived in the Einstein-de Sitter Universe approximation are recovered when the six coefficients are set equal to unity.

In future analyses a potential route to extend the validity regime of the theoretical model for the bispectrum would be to use the "Effective Field Theory of LSS" formalism \cite{Baumann:2010tm,Carrasco:2012cv,Carrasco:2013mua,Pajer:2013jj,Mercolli:2013bsa} for the bispectrum as done in \cite{Baldauf:2014qfa,Angulo:2014tfa}; recently  both real space bispectrum \cite{Steele:2020tak} and trispectrum \cite{Steele:2021lnz} models were calibrated in the EFT of LSS formalism at 1-loop order. For our present application, however, we would need the redshift-space counterpart of these results.
The redshift space kernels $Z^{(1)}$, $Z^{(2)}$ are obtained from the $F^{(2)}$ and $G^{(2)}$ kernels, see Equation~\ref{eq:zkernels} in Appendix \ref{app:fast_tk_model}. Hence the models for galaxy bispectrum and trispectrum are the tree-level-like expressions:
\begin{eqnarray}
\label{eq:bktk_grav}
B_\mr{FPT}(\mf{k}_1, \mf{k}_2, \mf{k}_3) &=&\, 2 
Z^{(1)}_\mr{FPT}\left[\mf{k}_1\right]
Z^{(1)}_\mr{FPT}\left[\mf{k}_2\right]
Z^{(2)}_\mr{FPT}\left[\mf{k}_1,\mf{k}_2\right]\,P\left(k_1\right)P\left(k_2\right)
\notag \\
&+& \quad 2 \,\,\mr{permutations}\,,
\notag \\
\notag \\
T_\mr{FPT}(\mf{k}_1,\mf{k}_2,\mf{k}_3,\mf{k}_4)
&=&
4
P(k_1)P(k_2)
Z^{(1)}_\mr{SPT}\left[\mf{k}_1\right]
Z^{(1)}_\mr{SPT}\left[\mf{k}_2\right]
\notag \\
&\times& \Big\{
Z^{(2)}_{\mr{FPT}}\left[\mf{k}_1,-\mf{k}_{13}\right]
Z^{(2)}_{\mr{FPT}}\left[\mf{k}_2, \mf{k}_{13}\right]
P(k_{13})
\notag \\
&+&
Z^{(2)}_{\mr{FPT}}\left[\mf{k}_1,-\mf{k}_{14}\right]
Z^{(2)}_{\mr{FPT}}\left[\mf{k}_2, \mf{k}_{14}\right]
P(k_{14})
\Big\}
\,+\,5\,\mr{p}.
\notag \\
&+&
6
\,\,
Z^{(1)}_\mr{SPT}\left[\mf{k}_1\right]
Z^{(1)}_\mr{SPT}\left[\mf{k}_2\right]
Z^{(1)}_\mr{SPT}\left[\mf{k}_3\right]
Z^{(3)}_{\mr{FPT}}\left[\mf{k}_{1},\mf{k}_2, \mf{k}_{3}\right]
P(k_1)P(k_2)P(k_3)
\notag \\
&+&\,3\,\mr{p}. ,
\end{eqnarray}
\noindent where in the above expression $P\equiv P_{\delta\delta}$ while "FPT/SPT" stands for fitted/standard perturbation theory kernels.

Detecting imprints of primordial non-Gaussianity (PNG) on the galaxy density field is one of the main purposes for including higher-order statistics in LSS analyses. In this work we consider the local type PNG signature present in both matter bispectrum and i-trispectrum, which is usually parameterised by $f_{\rm nl}$ e.g., \cite{Verde:1999ij,Komatsu:2001rj}. These are, at first order in $f_\mr{nl}$,
\begin{align}
\label{eq:png_bktk}
B^{\mr{PNG}}(\mf{k}_1,\mf{k}_2,\mf{k}_3) &= 
Z^{(1)}(k_1)Z^{(1)}(k_2)Z^{(1)}(k_3)
\dfrac{\mathcal{M}(k_1)}{\mathcal{M}(k_2)\mathcal{M}(k_3)}
\dfrac{2f_{\mr{nl}}}{c^2}
P(k_2)P(k_3) \;+\; \mr{cyc.}\,.
\notag\\
T^{\mr{PNG}}(\mf{k}_1,\mf{k}_2,\mf{k}_3,\mf{k}_4)
&=\,
\dfrac{f_{\mr{nl}}}{c^2}
Z^{(1)}(k_1)Z^{(1)}(k_2)Z^{(1)}(k_3)
\notag\\
&\times\,
\Bigg\{
\left[
4 
\dfrac{\mathcal{M}(k_1)}{\mathcal{M}(k_2)}
P(k_2)P(k_3)
\dfrac{P(|\mf{k}_3+\mf{k}_4|)}
{\mathcal{M}(|\mf{k}_3+\mf{k}_4|)}
Z^{(2)}_\mr{FPT}\left[-\mf{k}_3,\mf{k}_3+\mf{k}_4\right]\; + \;5\,\mr{p.}\right]
\notag\\
&+\,
\left[
2
\dfrac{\mathcal{M}(|\mf{k}_3+\mf{k}_4|)}{\mathcal{M}(k_1)\mathcal{M}(k_2)}
P(k_1)P(k_2)P(k_3)Z^{(2)}_\mr{FPT}\left[\mf{k}_3+\mf{k}_4,-\mf{k}_3\right]\;+\;2\,\mr{p.}
\right]
\Bigg\} \notag\\
\;&+\;3\,\mr{p.}\,.
\end{align}{}
\noindent where $\mathcal{M}_k=\frac{3}{5}\,k^2\mathbb{T}_k\,D_+/(\Omega_\mr{m}H_0^2)$, with $D_+$ being the growth factor and $\mathbb{T}_k$ the power spectrum transfer function.
For a more extensive description and derivation please see appendix E of \cite{Gualdi:2020eag}.

For simplicity, we do not include here the scale dependent bias effect induced by a local $f_\mathrm{NL}$ \cite{Dalal:2007cu,Matarrese:2008nc} which appears primarily in the power spectrum at large scales. Corresponding contributions arising from the multivariate bias expansion are present also in the bispectrum \cite{Giannantonio:2009ak,Baldauf:2010vn,Tellarini:2015faa} and trispectrum \cite{Lazeyras:2015giz}.


\subsection{Choice of model's parameters}
\label{sec:choice_params}
After calibrating the $F^{(2)}$ and $G^{(2)}$ kernels, their coefficients are kept fixed. The model includes cosmological and "nuisance" parameters. Assuming locality in Lagrangian space, tracers bias is parametrised at third order by $b_1$, $b_2$, $b_3$ \cite{Abidi:2018eyd}, the standard bias parameters. The tidal bias is given by $b_\mr{s} = -4/7(b_1-1)$ together with the third order non local bias being $b_{3,\mr{nl}} = (32/315)(b_1-1)$ \cite{Chan:2012jj,Baldauf:2012hs,Saito:2014qha}. For dark matter particles all these bias parameters are expected to be zero (beside $b_1=1$), however we keep them free to vary to emulate the parameter space considered when fitting biased tracers. $\sigma_P$, $\sigma_B$ and $\sigma_\mathcal{T}$ are the phenomenological fingers-of-God damping parameters \cite{Gualdi:2020eag}. Finally $A_\mr{n}$ parametrises deviations from Poissonian shot-noise as done for example in the BOSS analysis of the bispectrum monopole \cite{Gil-Marin:2016wya} (see Equation \ref{eq:an_param} ). The above seven parameters should be seen as nuisance parameters. 

Cosmological information is captured by $f$, $\sigma_8$, $\alpha_\parallel$, $\alpha_\perp$ and $f_\mr{nl}$.
$f$ is the growth rate defined as the logarithmic derivative of the growth factor with respect to the scale factor $d\ln{D_+}/d\ln{a}$. $f_\mr{nl}$ is the amplitude of local PNG whose imprint on bispectrum and i-trispectrum was described in Section \ref{sec:theo_models}.

We take into account the Alcock-Paczy\`{n}ski dilation effect \cite{Alcock:1979mp} through the scaling parameters $\alpha_\parallel = k_\parallel/p_\parallel$, $\alpha_\perp = k_\perp/p_\perp$, which relates the measured wave-vector $\mf{k}$ for a particular cosmology with the corresponding true wave-vector $\mf{p}$. These scaling parameters deviate from unity when the cosmology used to transform redshifts into co-moving distances is different from the true one. Such spurious anisotropy ($\alpha_\parallel/\alpha_\perp\neq1$) and isotropic change of scale ($\alpha_\parallel {\alpha_\perp}^2 \neq1$) allow us to set constraint the cosmological parameters when interpreted for a specific cosmological model. In this paper we always work with the measured true co-moving position, and therefore we always expect these scaling parameters to be unity, but we allow them to vary in the model, emulating the actual fit one should perform on real data. For simplicity we work with true underlying template and do not account for the usual horizon scale dependence within the $\alpha$'s definition, which would account for the relative shift of the BAO peak-position on the data with respect to the template used.

The reason for this choice, in particular $\alpha_\parallel$, $\alpha_\perp$ (and $f$) instead of $\Omega_m$, $h$,$w$ etc., is that in LSS clustering analyses a necessary step, even before measuring the statistics from the data, is to convert the object's measured redshifts into distances. To do so, a fiducial cosmology has to be assumed. After that, it is customary to adopt a fixed template, corresponding to that of the fiducial cosmology, for the real space matter power spectrum and bispectrum (and thus trispectrum). Hence the robust physical quantities that can be extracted from the data are the Alcock-Paczy\`{n}ski parameters, (and $f$ and $\sigma_8$), see e.g., \cite{Brieden:2020upf} and \cite{Brieden:2021edu}.


\subsection{Multipoles expansion}
\label{sec:multipoles_exp}
Before reporting the expression for the multipoles, recall that the Alcock-Paczy\`{n}ski effect 
also modifies the angles between the $k$-vectors and the line of sight. Given the cosines $\mu$ and $\eta$ of the angles between the line of sight and $\mf{k}$ and $\mf{p}$ respectively, the conversion is given by \cite{Gil-Marin:2016wya}

\begin{align}
  \label{eq:ap_par_rels}
  p &= |\mf{p}| = \dfrac{k}{\alpha_\perp}\left[1+\mu^2\left(F^{-2}-1\right)\right]^{\frac{1}{2}} 
  \notag \\
  \eta &= \dfrac{\mu}{F}\left[1+\mu^2\left(F^{-2}-1\right)\right]^{-\frac{1}{2}}\,,
\end{align}
\noindent where $F\equiv\alpha_\parallel / \alpha_\perp$. 

The same expansion in terms of Legendre polynomials, ${\cal L}_{\ell}$, is applied to power spectrum, bispectrum and i-trispectrum. Here we will only use $\ell=0$ for the monopole and $\ell=2$ for the quadrupole. We do not include the hexadecapole signal because the "discrete $\mu$" binning effect described in \cite{Beutler:2016arn} becomes dominant,  given the  size of the simulations boxes.
The Legendre polynomials are computed as a function of the cosine of the angle between a given $k$-vector and the line of sight. We thus obtain, 
\begin{eqnarray}
\label{eq:pk_model}
P^{(\ell)}(k) = \dfrac{(2\ell + 1)}{2\alpha_\parallel\alpha_\perp^2}\int^{+1}_{-1}d\mu \,\mathcal{L}_\ell(\mu)\,P(p,\eta)\,
\end{eqnarray}{}
\begin{eqnarray}
\label{eq:bk_model}
B^{(\ell_i)}(k_1,k_2,k_3) = \dfrac{(2\ell+1)}{8\pi \alpha_\parallel^2\alpha_\perp^4}\int^{+1}_{-1}d\mu_1\int^{2\pi}_0 d\phi \,
\mathcal{L}_\ell(\mu_i)B(p_1,p_2,p_3,\eta_1,\eta_2)\,,
\end{eqnarray}{}
\noindent
where $\mu_i$ with $i=1,2,3$ is the cosine of the angle between the $i$-th $k$-vector and the line of sight. 
In the above equation, we choose to use $\mu_1$ and $\phi$ to integrate over all the possible triangle orientations with respect to the line of sight.
The azimuthal angle $\phi$ around $\mf{k}_1$,
is defined  by $\mu_2\equiv\mu_1\cos\theta_{12} - \sqrt{1-\mu_1^2}\sqrt{1-\cos\theta_{12}^2}\cos\phi$ (see~\cite{Scoccimarro:1999ed}).

Equation~\ref{eq:bk_model} is used to compute the bispectrum multipoles such as $B^{(\ell_1)}\equiv B^{(\ell00)}$, $B^{(\ell_2)}\equiv B^{(0\ell0)}$ and $B^{(\ell_3)}\equiv B^{(00\ell)}$.
When all possible multipoles of the same order $\ell$ are included in the data-vector, the shorter $B^{(0,\ell)}$ notation is used (monopole plus all the $\ell$-multipoles).

Similarly to the power spectrum case, and for the same reasons,  also for the bispectrum we only use the quadrupole to probe the anisotropic signal.
Nevertheless in future analyses and for larger volumes, significant additional information can be harvested if higher multipoles are included in the data-vector, as we have recently shown in \cite{Gualdi:2020ymf}.

Analogously, and following closely \cite{Gualdi:2020eag}, we write the i-trispectrum multipole expansion as:
\begin{eqnarray}
\label{eq:tk_model}
\mathcal{T}^{(\ell_i)}\left(k_1,k_2,k_3,k_4\right)
&&=\,
\dfrac{1}{3}\sum_{\substack{k_1,k_2,k_3,k_4 \\ k_1,k_3,k_2,k_4 \\ k_1,k_2,k_4,k_3}}
\dfrac{2\ell + 1}{16\pi^2 \Delta D \alpha_\parallel^3\alpha_\perp^6}
\int^{D_\mr{max}}_{D_\mr{min}}dD\int^{+1}_{-1}\,d\mu_D\int^{2\pi}_0\,d\phi_{12}\int^{2\pi}_0\,d\psi
\notag \\
&&\times\,
\mathcal{L}_\ell(\mu_i)
T^\mr{s}\left(p_1,p_2,p_3,p_4,D,\eta_D,\phi_{12},\psi\right)
\,.
\end{eqnarray}

\noindent where also for the i-trispectrum multipoles expansion $i=1,2,3,4$ and $i$ highlights which angle between a particular $k$-vector and the line of sight has been used, giving for example $\mathcal{T}^{(\ell_1)}\equiv\mathcal{T}^{(\ell000)}$, $\mathcal{T}^{(\ell_2)}\equiv\mathcal{T}^{(0\ell00)}$, $\mathcal{T}^{(\ell_3)}\equiv\mathcal{T}^{(00\ell0)}$ and $\mathcal{T}^{(\ell_4)}\equiv\mathcal{T}^{(000\ell)}$.
As for the bispectrum, also for the i-trispectrum the short notation $\mathcal{T}^{(0,\ell)}$ is used to indicate the monopole plus all the multipoles of order $\ell$.

$D$ is one of the two diagonals of the quadrilaterals, defined by $\mf{D}+\mf{k}_1+\mf{k}_2=0$ or $\mf{D}-\mf{k}_3-\mf{k}_4=0$.
The angle $\psi$ describes the "folding" of the quadrilateral around the diagonal $\mf{D}$, in other words, the angle between the two planes defined by the two triangles $\mf{D}+\mf{k}_1+\mf{k}_2=0$ and $\mf{D}-\mf{k}_3-\mf{k}_4=0$.
The angle $\phi_{12}$ determines the rotation of the triangle $\mf{D}+\mf{k}_1+\mf{k}_2=0$ away from the $xz$-plane (see Figure 1 of \cite{Gualdi:2020eag}). 
Finally $\mu_D$ is the cosine of the angle between the diagonal, $\mf{D}$, and the line of sight. By varying $(\psi,\phi_{12},\mu_D)$, it is possible to span all the possible orientations of the quadrilateral with respect to the line of sight.
Equation \ref{eq:tk_model} is the natural extension beyond the monopole of Equation 2.7 in \cite{Gualdi:2020eag}.

The multi-dimensional integral in Equation \ref{eq:tk_model} requires a computational time for the i-trispectrum data-vector which is too long to be suitable for parameter inference through MCMC sampling. To overcome this obstacle, in Appendix~\ref{app:fast_tk_model} we present a decomposition of the integrand such that the dependence on certain combinations of cosmological parameters can be factorised out of the integrals. We then pre-compute the individual multi-dimensional integrals corresponding to each of these terms on a three-dimensional grid, sampling the relevant range for the remaining parameters that cannot be factorised: $(\alpha_\parallel,\alpha_\perp,\sigma_\mathcal{T})$.

This stratagem (which we call  $\mathcal{T}$-Expreso) allows us to instantly evaluate the i-trispectrum model (approximately $\sim 2800$ times faster than computing $\mathcal{T}^{(0,2)}$ with the standard approach) for arbitrary values of the remaining parameters $(b_1, b_2, b_3, f, \sigma_8, f_\mr{nl}, A_\mr{n})$. More details are given in Appendix~\ref{app:fast_tk_model}.

To model the shot-noise contribution, we proceed similarly to what done in \cite{Gualdi:2020eag} and account for deviations from Poissonian statistics by including a free amplitude parameter $A_\mr{n}$ which is the same for all the statistics, as done in the BOSS analyses \cite{Gil-Marin:2014pva,Gil-Marin:2016wya}. More details and tests on the accuracy of this prescription are presented in Appendix \ref{sec:app_shot_noise}.

\subsection{Estimators: measuring P, B, \texorpdfstring{$\mf{\mathcal{T}}$}{}}
\label{sec:estim}

In order to measure the signal from synthetic data in cubic boxes, we use the same estimators and procedure presented in \cite{Gualdi:2020eag}.
Given a cubic box filled with a distribution of dark matter particles, the first step consist in applying redshift space distortions to each particle by shifting its position along the chosen line of sight ($\mf{z}$-axis in our case) by 

\begin{eqnarray}
r_\mr{z'} = r_\mr{z} + \dfrac{v_\mr{z}}{H(z)}(1+z),
\end{eqnarray}

\noindent where $z$ is the particle's redshift and $H$ is the Hubble parameter evaluated at $z$.

The overdensity field in configuration space is then obtained by a mass assignment procedure \cite{2010PhDT.........4J,Sefusatti:2015aex} which distributes each particle's contribution onto a three-dimensional discrete grid.
After this step the corresponding overdensity field in Fourier space is computed through discrete Fast Fourier Transform algorithms (FFT) \cite{10.1145/1464291.1464352}.

The estimators for power spectrum, bispectrum and i-trispectrum are based on the ones introduced by \cite{Scoccimarro:2000sn,Scoccimarro:2015bla} which were also used in \cite{Tomlinson:2019bjx}. In particular the performance of the trispectrum estimator is discussed in details in \cite{Gualdi:2020eag}. 
The only additional step required for this analysis was to modify the quantity defined in equation (2.9) of \cite{Gualdi:2020eag} by adding the Legendre polynomials to estimate the multipoles of the over-density field. This, which is the inverse Fourier transform of the density field over a shell of radius $k$ and thickness $\Delta k$ for each 3D pixel of the simulation box, becomes
\begin{eqnarray}
\label{eq:IJ_def_maintext}
I_{k}(\mathbf{x}) = \int_{k}\dfrac{d\mathbf{q}^3}{(2\pi)^3}\,\delta_{\mathbf{q}}
e^{-i\mathbf{x}\mathbf{q}}
\Longrightarrow
I_{k}^{(\ell)}(\mathbf{x}) = \int_{k}\dfrac{d\mathbf{q}^3}{(2\pi)^3}\,\mathcal{L}_\ell(\mu)\,\delta_{\mathbf{q}}
e^{-i\mathbf{x}\mathbf{q}}\,,
\end{eqnarray}

\noindent with $\mu$ being the cosine of the angle between the wave-vector $\mf{q}$ and the line of sight.

For both power spectrum and bispectrum multipoles, the shot-noise is estimated from  each individual catalogue and subtracted from the statistics measurements. For the i-trispectrum we found less noisy to compute  analytically the shot-noise term and  subtract it afterwords  from each catalogue measurements. More details are given in Appendix \ref{sec:app_shot_noise}.

\subsection{Signal-to-noise ratio and Fisher formalism}
\label{sec:sn_an_fisher}
The standard procedure to derive parameter constraints involves sampling the multidimensional parameter posterior distributions through Markov chain Monte Carlo (MCMC) algorithms. 
In this application the full data-vector is composed of the power spectrum monopole and quadrupole, the bispectrum monopole and quadrupoles and the trispectrum monopole and quadrupoles, and we are interested in estimating the additional statisical power offered by the less explored statistics i.e., the trispectrum and/or bispectrum multipoles. 

A full exploration with MCMC is not straightforward or fast: the likelihood evaluation gets increasingly computationally expensive as higher-order correlators are included in the data-vector. 
For this reason we find useful to obtain a preliminary (approximate) estimate of the additional constraining power provided by the new terms of data-vector by looking at the cumulative signal-to-noise ratio (S/N).

The average (S/N) is computed as a function of the maximum $k$-value, using the data-vector measurements ($\mf{d}_i$) from each realisation (indicated by the index $i$) and the corresponding covariance matrix $\mf{Cov}_\mf{d}$ estimated from the whole set of $N_{\rm sim}$ simulations (see Section~\ref{sec:analysis_setup} below for more details on estimation of the covariance matrix):
\begin{eqnarray}
\label{eq:signal_noise}
\langle\mr{S/N}\rangle\quad = 
\quad\dfrac{1}{N_\mr{sim}}\sum_{i=1}^{N_\mr{sim}}\sqrt{ \hat{\mf{d}}^\intercal_i\;\mf{Cov}_\mf{d}^{-1} \; \hat{\mf{d}}_i}\,,
\end{eqnarray}
\noindent where the measured data-vectors $\hat{\mf{d}}$ has a varying cut-off scale $k_{\rm max}$.

Given a theoretical model description of the data-vector, ${\bf d}_{\rm th}(\theta_i)$, with dependence on model parameters $\theta_i$, the Fisher information matrix is the Hessian matrix of the associated log likelihood, $\mathcal{L}$, with the derivatives taken with respect to the model parameters,
\begin{eqnarray}
F_{ij}\quad=\quad -\Bigg\langle\dfrac{\partial^2\mathcal{L}}{\partial\theta_i\partial\theta_j} \Bigg\rangle\Bigg|_{\theta_\mr{ml}}\,,
\end{eqnarray}
\noindent where $\theta_\mr{ml}$ indicates that the derivatives are evaluated at the likelihood maximum.

When estimating the minimum errors for each parameter in a multi-parameter case analysis, this is given by the square root of the corresponding Fisher information inverse matrix element $\Delta\theta^i_\mr{min}=\sqrt{F_{ii}^{-1}}$ \cite{Tegmark:1996bz}.

Using the data-vector's covariance matrix and its derivatives with respect to the model parameters, each element of the Fisher information matrix can be computed as
\begin{eqnarray}
\label{eq:fish_info}
F_{ij} = \dfrac{\partial \mf{d}_{\rm th}^\intercal}{\partial\theta_i}\,\mf{Cov}_\mf{d}^{-1}\,\dfrac{\partial \mf{d}_{\rm th}}{\partial\theta_j}\,,
\end{eqnarray}

\noindent since here (as in most Fisher forecasts and in most LSS studies) we adopt a Gaussian likelihood with fixed covariance matrix (see e.g., \cite{Carron:2012pw,Kalus:2015lna}).
For a recent Fisher forecast study using data-vector's derivatives estimated from measurements on synthetic data  exploring the additional constraining power provided by the bispectrum monopole, especially regarding neutrino mass constraints see \cite{Hahn:2020lou}.

The Fisher matrix estimate of parameters errors is well known to be an approximation, but we also find it useful to explore a variety of data-vector choices which would be prohibitive or requiring much more effort to explore through MCMC sampling.

\subsection{Analysis set-up}
\label{sec:analysis_setup}

We consider different combinations of the full data-vector including multipoles of the power spectrum, bispectrum and i-trispectrum:
\newline
$(P^{(0)},P^{(2)},B^{(0)},B^{(200)},B^{(020)},B^{(002)},\mathcal{T}^{(0)},\mathcal{T}^{(2000)},\mathcal{T}^{(0200)},\mathcal{T}^{(0020)},\mathcal{T}^{(0002)})$ which in short notation is labelled as $(P^{(0,2)},B^{(0,2)},\mathcal{T}^{(0,2)})$.

The improvements offered by the less explored statistics (bispectrum and trispectrum multipoles) on the constraints for the set of parameters $(b_1,b_2,b_3,f,\sigma_8,f_\mr{nl},\alpha_\parallel,\alpha_\perp,\sigma_P,\sigma_B,\sigma_\mathcal{T},A_\mr{n})$ is the target of our analysis.
In terms of scale range we use

\begin{itemize}
  \item power spectrum: $k_\mr{min}(P^{(0)})=0.04\, h\rm{Mpc}^{-1}$, $k_\mr{min}(P^{(2)})=0.06\, h\rm{Mpc}^{-1}$ \newline and $k_\mr{max}(P^{(0,2)})~=~0.13\, h\rm{Mpc}^{-1}$;
  \item bispectrum: $k_\mr{min}(B^{(0,2)})=0.04\, h\rm{Mpc}^{-1}$ and $k_\mr{max}(B^{(0,2)})=0.12\, h\rm{Mpc}^{-1}$;
  \item i-trispectrum: $k_\mr{min}(\mathcal{T}^{(0,2)})=0.04$ $h\rm{Mpc}^{-1}$ and $k_\mr{max}(\mathcal{T}^{(0,2)})=0.12$ $h\rm{Mpc}^{-1}$;
\end{itemize}
\noindent For both bispectrum and i-trispectrum we choose a $k_\mr{max}=0.12\,h\rm{Mpc}^{-1}$ as a compromise between minimum scale necessary to constrain the fingers-of-God parameters $(\sigma_B,\sigma_\mathcal{T})$ and the theoretical model ability to well fit the data.
With these settings, the power spectrum, bispectrum and i-trispectrum data-vector reaches a dimension of 3615 elements.

 The box is discretised in $256^3$ grid cells, and we consider a bin size of $\Delta k=1.1\times k_\mr{f}$, where $k_\mr{f}=0.00625\, h\rm{Mpc}^{-1}$ is the fundamental frequency of the simulations boxes.

Each simulation is randomly sub-sampled down to a number density of $n_\mr{p}=5\times 10^{-4}$ $(h\mr{Gpc}^{-1})^{-3}$, to be in a realistic density-regime scenario. Having the fully sampled and sub-sampled density for each realisation allows us to test the shot-noise subtraction procedure on each catalogue. 

We account for shot-noise (allowing for deviations from Poisson statistics through the free parameter $A_{\rm n}$) by subtracting it from the measured signal,  as described in Appendix \ref{sec:app_shot_noise}.  

The covariance matrix is estimated from 8000 realisations, which is not significantly larger than the size of the full data-vector, yielding well known shortcomings \cite{Hartlap:2006kj}. 
To account for this we proceed as follows.
When looking at the signal-to-noise ratio and performing Fisher forecasts, where, as described in Section \ref{sec:sn_an_fisher}, the relevant quantity is an estimate of the 1D $68\%$ credible region from the inverse of the covariance matrix, we apply the Hartlap factor correction \cite{Hartlap:2006kj}. We have checked that the accuracy of the above correction is sufficient for the purposes of this work when resorting to the (S/N) and Fisher forecasts. 
On the other hand, when sampling the parameters posterior through MCMC, we adopt the Sellentin and Heavens prescription to modify the likelihood \cite{Sellentin:2015waz}. This correction is indeed more accurate especially in the tails of the posterior distribution. Both approaches are described in Appendix \ref{app:sel_heav_corr}.

The obtained covariance matrix is suitably re-scaled (divided by a factor of $25$) in order to simulate a survey volume of $25$ $(h^{-1}\mr{Gpc})^3$. With this choice of volume,  redshift ($z=0.5$), and the above settings for the shot-noise we realistically match the specs of current and near-future spectroscopic clustering data-sets.

When running MCMC sampling we assume a Gaussian likelihood together with employing flat uninformative priors on the constrained parameters. In particular the prior ranges for the different parameters are $b_1:[0.7, 1.3]$ , $b_2: [-0.5, 0.5]$, $b_3: [-0.8, 0.8]$, $f: [0.5, 1.]$, $\sigma_8: [0.5, 1.]$, $\alpha_\parallel: [0.96, 1.04]$, $\alpha_\perp: [0.96, 1.04]$, $\sigma_P: [2., 7.]$, $\sigma_B: [2., 9.]$, $\sigma_\mathcal{T}: [3.5,6.5]$,
    $f_\mr{nl}: [-1000, 1000]$ and  $A_\mr{n}: [-0.6, 0.6]$.

It is important to note that we neglect here the effect that the PNG parameter $f_{\rm nl}$ has on the scale dependent halo bias~\cite{Matarrese:2008nc,Dalal:2007cu}. It is well known that the non-Gaussian halo bias offers a very promising avenue to constrain $f_{\rm nl}$ from the measurement of the power spectrum of biased tracers. However, not only it cannot be tested on the available simulations (as they are dark matter only and we do not consider halo catalogs) but also it exploits a different physical mechanism and requires a good understanding of the tracer's bias behaviour as function of mass (see e.g.,~\cite{Reid:2010vc}). In future applications it is reasonable (and wise) to keep the pure clustering analysis and the halo bias analysis of $f_{\rm nl}$ separated as to offer a consistency check and only combine the constraints on the parameter later. For this reason we leave the inclusion of non-Gaussian halo bias to future work. In what follows, when comparing the $f_{\rm nl}$ constraints obtained for the extended and baseline data-vectors this caveat should be kept in mind. 

\section{Results}
\label{sec:results}

\begin{figure}[tbp]
\centering 
\includegraphics[width=0.98\textwidth]
{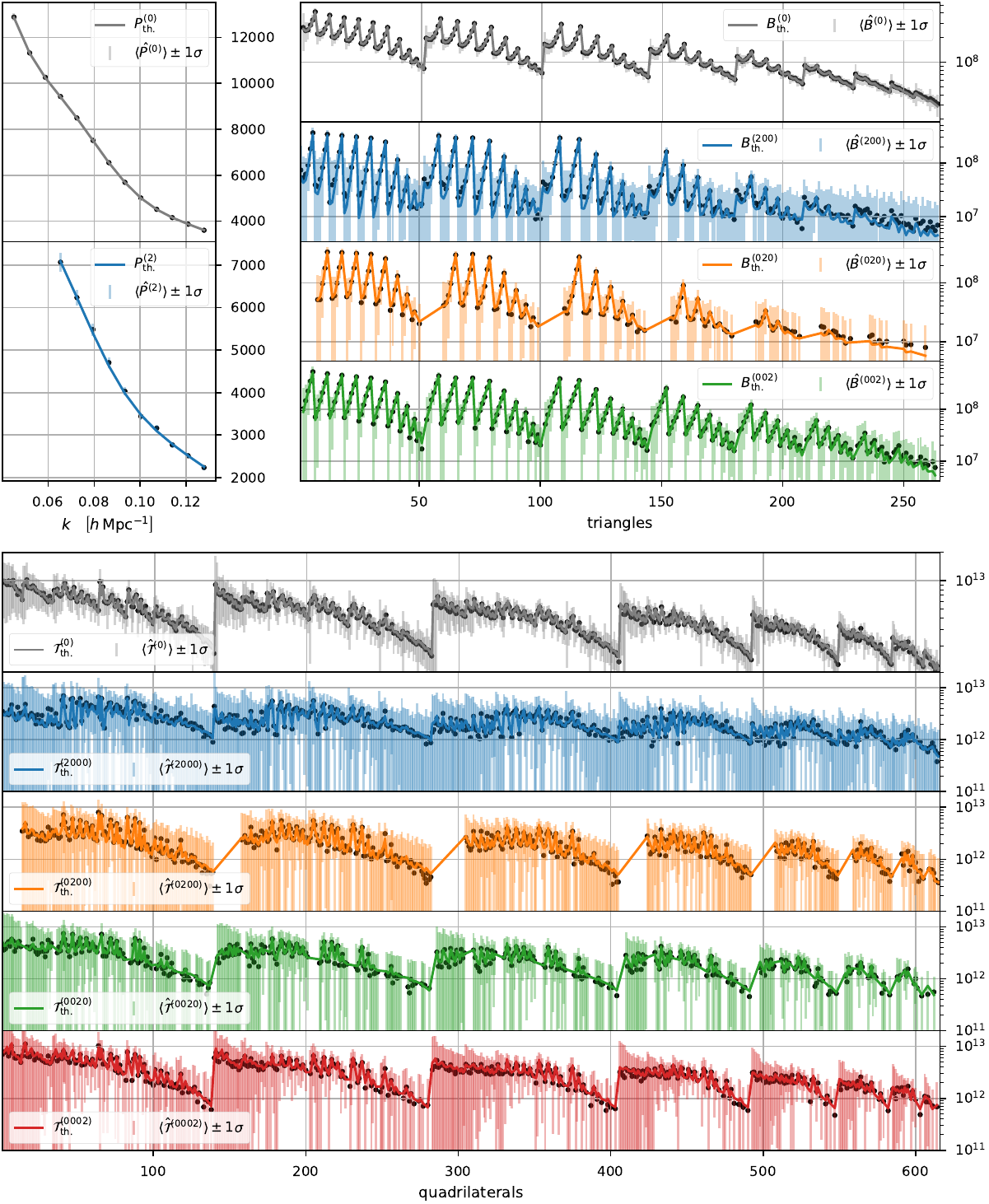}
\caption{\label{fig:fit_pbt_stats} 
Measurements (mean of the 8000 simulations, points), $1\sigma$ error ( rescaled to simulate a single realisation with observed volume equal to 25 $(h^{-1}\mr{Gpc})^3$) and theoretical models (continuous lines) for the monopoles and quadrupoles terms of power spectrum, bispectrum and i-trispectrum. In the quadrupoles of both bispectrum and i-trispectrum, some configurations are omitted to avoid redundancy (see text for more details). The triangles configuration in the x-axis are ordered as in \cite{Gualdi:2020ymf} and the skew-quadrilaterals as in \cite{Gualdi:2020eag}. The theoretical model has been computed using the best-fit parameters derived from the analysis in Section \ref{sec:mcmc_sampling}.}
\end{figure}

\begin{figure}[tbp]
\centering 
\includegraphics[width=0.985\textwidth]
{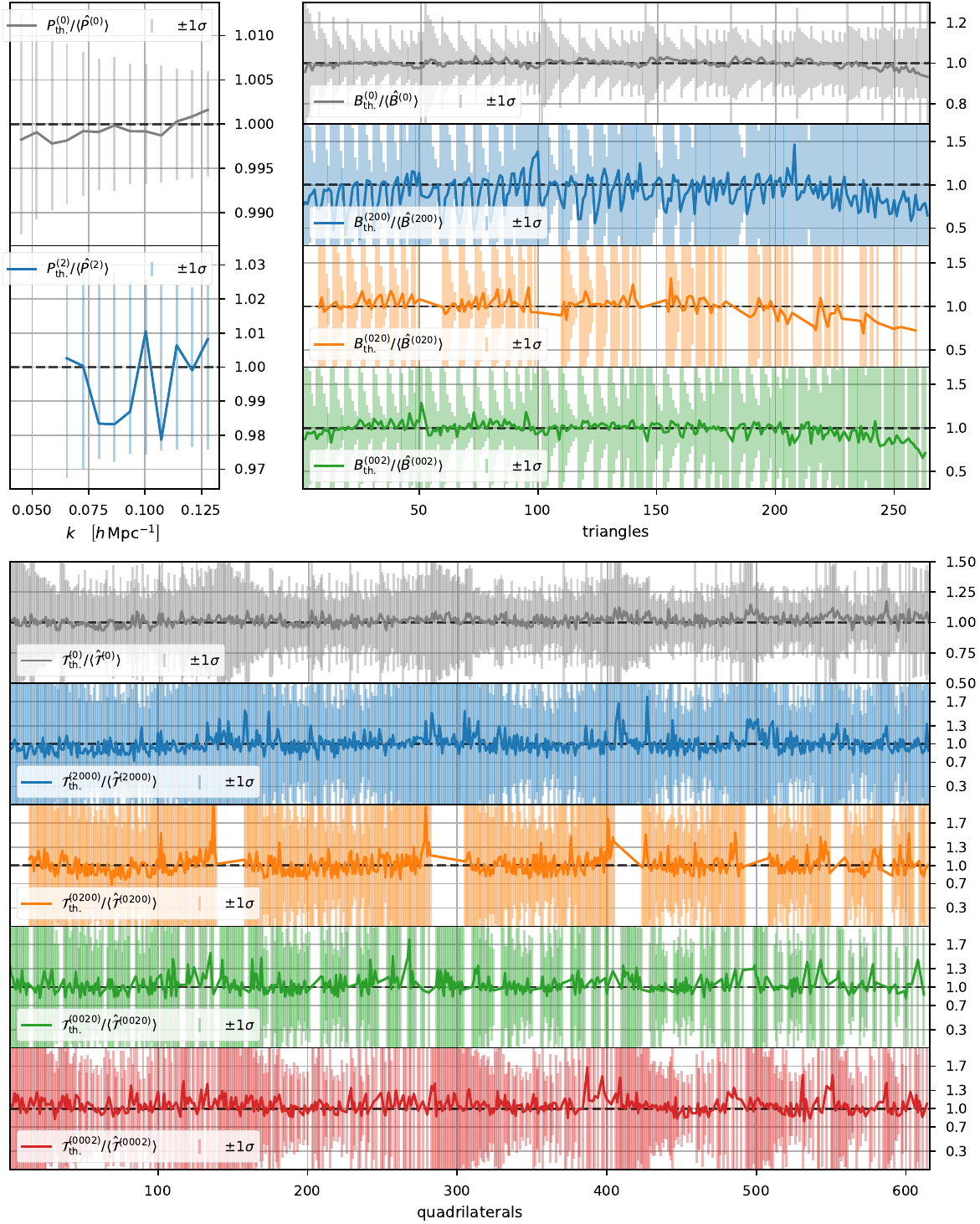}
\caption{\label{fig:fit_pbt_ratios} 
Same as Figure \ref{fig:fit_pbt_stats} but showing the ratios between the theoretical model and the mean of the measurements for each statistics. The errorbars are rescaled to simulate a single realisation with volume 25 $(h^{-1}\mr{Gpc})^3$.
The black dashed line corresponds to unity for each ratio.
}
\end{figure}

\subsection{Performance of the analytical model in describing the simulated data}
\label{sec:model_fit}
Before investigating the additional information content captured by the anisotropic redshift space bispectrum and i-trispectrum, we check the performance of the analytical model in fitting the data up to the chosen $k_\mr{max}$.
As measured data-vector we consider the mean of the complete set of 8000 simulations' measurements. This significantly reduces the cosmic variance uncertainty (the corresponding volume being 8000 $(h^{-1}\mr{Gpc})^3$).
Figure \ref{fig:fit_pbt_stats} shows for each statistics the theoretical model computed for the set of parameters corresponding to the best-fit values obtained in the analysis of Section \ref{sec:mcmc_sampling} --see sec \ref{sec:mcmc_sampling} for details-- and the measured data-vector. 

The error-bars correspond to a survey volume of $V_{\rm survey}=25\,(h^{-1}\mr{Gpc})^3$.
Even though the measured data-vector is the average of 8000 simulations, we re-scale the covariance only by a factor of 25 (recall that each simulations has a volume of 1 $(h^{-1}\mr{Gpc})^3$). This is a much larger error than the one corresponding to the mean of 8000 simulations, but it is of the same order of magnitude to what expected from current and future galaxy clustering surveys.
For these settings, the largest data-vector's fit including monopoles and quadrupoles for all the statistics, returns a $\chi^2\simeq1295$ for $N_\mr{dof}=3612 - 13$ (12 parameters plus using the mean of the realisation), which correspond to a reduced $\chi^2_\mr{red}\simeq0.36$. If we would have used as effective volume the one expected for the DESI survey (Table \ref{tab:surveys_specs}) $V_{\rm survey}=65\,(h^{-1}\mr{Gpc})^3$, the resulting reduced $\chi^2$ would have been $\chi^2_\mr{red}\simeq0.94$.
Analogously, Figure \ref{fig:fit_pbt_ratios} shows the ratios between the model and measurement of each statistic. 

Even with sub-percent relative error-bars, the power spectrum monopole model fits the data up to the adopted $k_\mr{max}$. The same is true for the power spectrum quadrupole. All the bispectrum multipoles show a good fit which starts worsening when reaching the maximum $k$, as it can be clearly seen in the top right rows of Figure \ref{fig:fit_pbt_ratios}. The i-trispectrum model accurately describes the measured signal for both monopole and quadrupole terms.

In both Figures~\ref{fig:fit_pbt_stats} and \ref{fig:fit_pbt_ratios} some configurations are omitted in the bispectrum and i-trispectrum quadrupoles data-vectors. For example to avoid redundancy, the bispectrum quadrupole for the triangle $(k_a,k_a,k_b)$ will be considered only in $B^{(200)}$ and $B^{(002)}$, not in $B^{(020)}$. Similarly for the i-trispectrum a quadrilateral $(k_a,k_a,k_b,k_c)$ will appear in $\mathcal{T}^{(2000)}$,$\mathcal{T}^{(0020)}$ and $\mathcal{T}^{(0002)}$ but not in $\mathcal{T}^{(0200)}$.

In the i-trispectrum data-vector some quadrilateral configurations have been excluded {\it a-priori} because after a first iteration of the kernel fitting procedure described in Appendix~\ref{sec:app_fit_ker}, the difference between model and mean measurement was uncomfortably high (above $15\%$). The form of the fitting function adopted can not yield an unbiased fit to these specific configurations. In practice this selection excludes 125 (out of 740) quadrilateral configurations in the i-trispectrum data-vector.
Furthermore all the symmetric quadrilateral configurations of the kind $(k_a,k_a,k_b,k_b)$ and $(k_a,k_a,k_a,k_a)$ have not been considered to avoid the (dominant)  contribution of unconnected terms while also limiting the dimension of the data-vector. 
The potential loss of constraining power resulting from this selection highlights the importance of improving the i-trispectrum theoretical modelling for future applications.

\begin{figure}[tbp]
\centering 
\includegraphics[width=1.0\textwidth]
{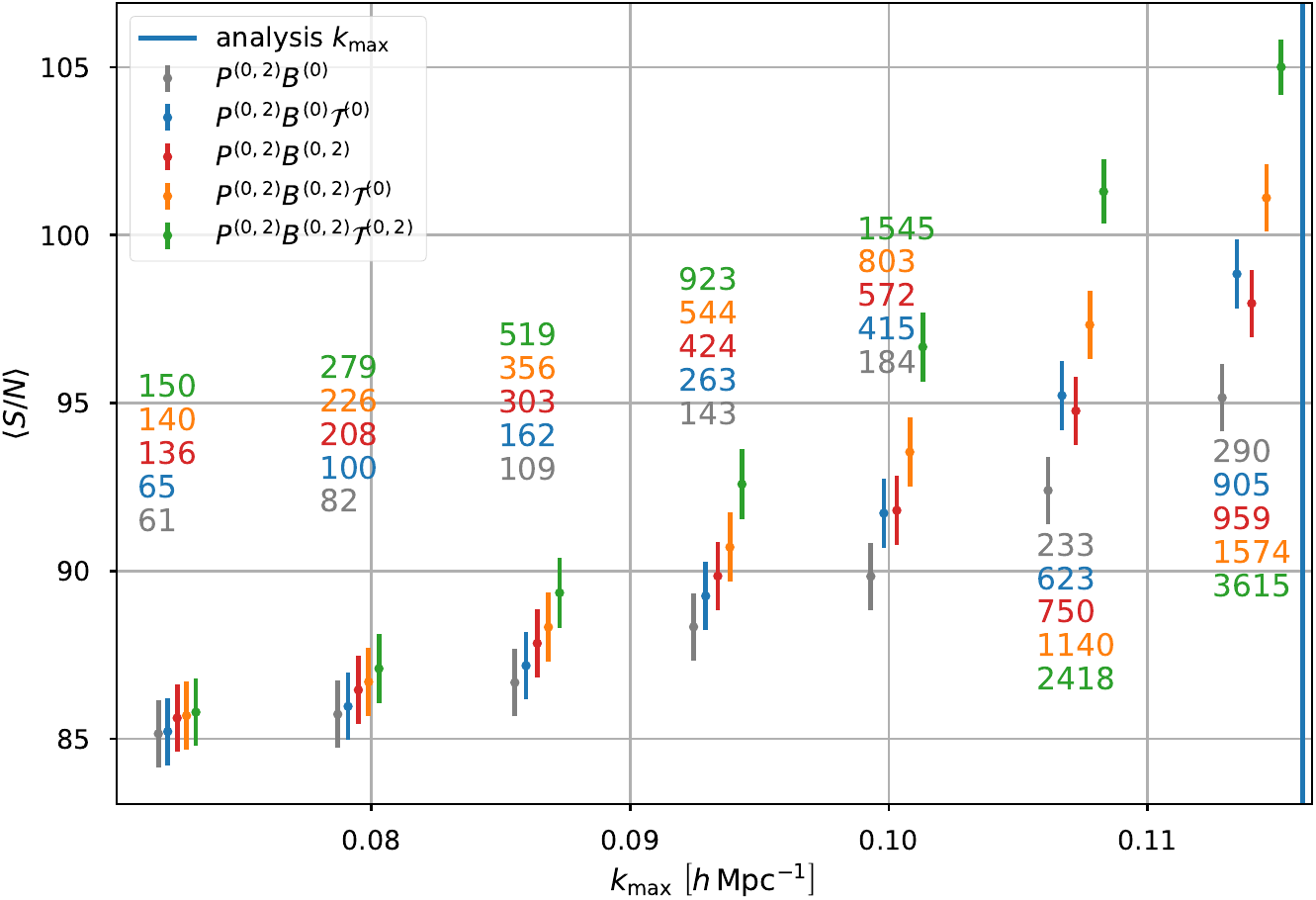}
\caption{\label{fig:signal_to_noise}
The average cumulative signal-to-noise ratio, $\langle \mr{S/N}\rangle$, for data-vectors corresponding to different combinations of the relevant statistics is shown as a function of $k_{\rm max}$, together with the standard deviation of $\langle \mr{S/N}\rangle$ estimated from the 8000 simulations. The (S/N) corresponds to that expected from realistic choices of effective volume and shot-noise, see Section~\ref{sec:analysis_setup}. Seven $k_{\rm max}$ values are represented: $k_{\rm max}=\,\left(0.0725, \,0.0795,\,0.0864,\, 0.0934,\, 0.1003 ,\, 0.1072 ,\, 0.1140\right)\,h\,\mr{Mpc}^{-1} $. In the plot, for clarity points are artificially slightly shifted from their exact $k$-value.
This figure shows that including up to all the i-trispectrum quadrupole terms significantly increases the $\langle \mr{S/N}\rangle$, with the gain becoming larger for greater values of $k_\mr{max}$.
It is very interesting to notice that the addition of the trispectrum monopole to the data-vector $\left[P^{(0,2)}B^{(0)}\right]$ yields an increment of $\langle \mr{S/N}\rangle$ comparable or larger than the one obtained by including all the bispectrum quadrupoles terms. Moreover, when considering the data-vector $\left[P^{(0,2)}B^{(0,2)}\mathcal{T}^{(0)}\right]$, the complementarity between the information captured by the trispectrum monopole and bispectrum quadrupoles becomes evident.
}
\end{figure}

\subsection{Preliminary results: (S/N) and Fisher forecasts}
\label{sec:snratio_res}

From the set of measurements on the simulations, it is straightforward to compute the cumulative signal-to-noise ratio as a function of $k_\mr{max}$ for the different statistics combinations without requiring any analytical modelling. This is displayed in Figure \ref{fig:signal_to_noise} where we use equation \ref{eq:signal_noise} for the different statistics (with the adjustments described in Section~\ref{sec:analysis_setup} to match realistic effective volume and noise).
The average signal to noise, $\langle \mr{S/N}\rangle$, from the simulation set is shown up to $k_{\rm max}=0.114 \,h/\mr{Mpc}$.
The error-bar is the corresponding (S/N) {\it rms} derived from computing the (S/N) for each of the 8000 simulations.

From Figure \ref{fig:signal_to_noise} it is clear that both bispectrum quadrupoles and i-trispectrum (monopole and quadrupoles) increase the total signal, with a greater effect as $k_\mr{max}$ increases. Moreover these two terms of the full data-vector prove to be complementary: by looking at the joint $(P^{(0,2)},B^{(0,2)},\mathcal{T}^{(0)})$ data-vector's signal on can appreciate that its corresponding (S/N) is larger than both the (S/N) ratios for $(P^{(0,2)},B^{(0,2)})$ and $(P^{(0,2)},B^{(0)},\mathcal{T}^{(0)})$. 
Moreover, Figure \ref{fig:signal_to_noise} shows that the i-trispectrum quadrupoles contain additional information with respect to the other statistics, motivating the effort done in this work to both measure and model this signal.
As described in Section \ref{sec:analysis_setup}, these results are for up to a minimum scale with $k_\mathrm{max}=0.12\,h\,\mr{Mpc}^{-1}$ at $z=0.5$, including modes only up to mildly non-linear regime where our model is still able to well describe the features of the different statistics as reported in Section \ref{sec:model_fit} and shown in Figures \ref{fig:fit_pbt_stats} and \ref{fig:fit_pbt_ratios}.
The aim of ongoing and future work is to extend the bispectrum  \cite{GilMarin:2011ik,Gil-Marin:2014pva,Hashimoto:2017klo,Bose:2018zpk,Steele:2020tak,Eggemeier:2021cam} and i-trispectrum \cite{Steele:2021lnz} modelling up to more non-linear scales, as it is done for example in the case of the power spectrum by the TNS model \cite{Taruya:2010mx}.

\begin{figure}[tbp]
\centering 
\includegraphics[width=1.0\textwidth]
{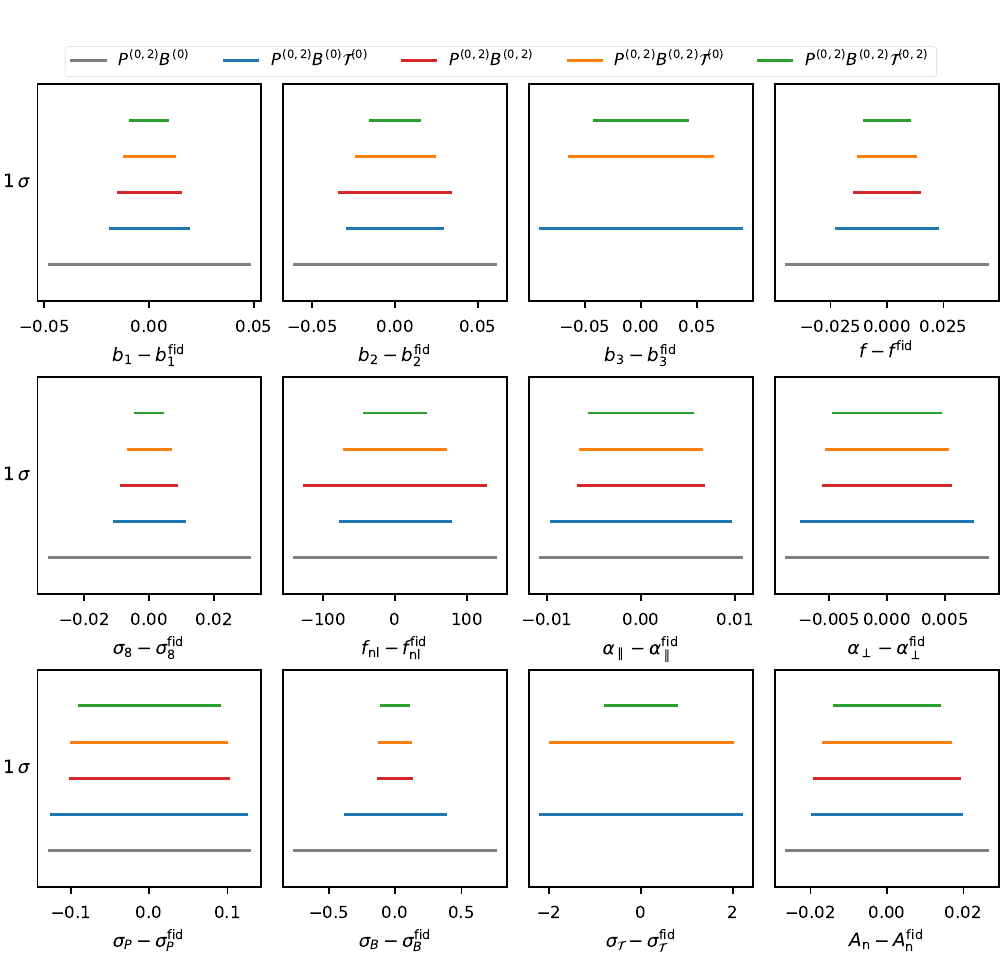}
\caption{\label{fig:fisher_forecasts} 
Fisher forecasts assuming a Gaussian likelihood and deriving the 1D 68$\%$ credible regions from equation \ref{eq:fish_info}. The derivatives of the data-vector with respect to the model parameters have been numerically computed from the theoretical model. The fiducial value of each parameter has been subtracted to center the error-bars around zero. The 1D 68$\%$ credible regions forecasted values are reported in Tables \ref{tab:table_constr}, \ref{tab:table_constr_nuisance} for the cosmological and nuisance parameters, respectively. 
}
\end{figure}

A first estimate of how this additional signal, encoded into both bispectrum and i-trispectrum multipoles, translates in terms of parameter constraints, can be studied through the Fisher forecasts formalism. 

Using the analytical models given in equations \ref{eq:pk_model}, \ref{eq:pk_model} and \ref{eq:tk_model} to numerically compute the derivatives for the both the set of nuisance $(b_1,b_2,b_3,\sigma_P,\sigma_B,\sigma_\mathcal{T},A_\mr{n})$ and cosmological $(f,\sigma_8,f_\mr{nl},\alpha_\parallel,\alpha_\perp)$ parameters, the 1D 68$\%$ credible regions can be forecasted using equation \ref{eq:fish_info}. These are displayed in Figure \ref{fig:fisher_forecasts}. The marginalised 1$\sigma$ parameter constraints convey a message in good agreement with what first observed in terms of signal-to-noise ratio. In addition to the complementarity between bispectrum and i-trispectrum, the Fisher forecasts highlight the impact of the i-trispectrum in improving the constraints on the primordial non-Gaussianity amplitude parameter $f_\mr{nl}$, similarly to what already reported in \cite{Gualdi:2020eag} but with the substantial difference of considering a much larger (and therefore more degenerate) parameter space. Even if to a smaller degree, also the quadrupole of the bispectrum is  sensitive to the primordial non-Gaussianity amplitude parameter:  as previously discussed in \cite{Gualdi:2020ymf}, at  the bispectrum level, redshift space distortions have different functional forms for the primordial and gravitational terms.

The constraints for the bias parameters $b_1$, $b_2$ and $b_3$ benefit in a similar fashion from all the terms of the joint data-vector.
For the cosmological parameters $f$ and $\sigma_8$ both i-trispectrum monopole and bispectrum quadrupoles tighten the marginalised 1D $68\%$ credible intervals when added to the data-vector, proving their complementarity.
The anisotropic bispectrum seems to be the most sensitive statistics to variation of the Alcock-Paczy\`{n}ski parameters $\alpha_\parallel$ and $\alpha_\perp$.

The forecasted 1D $68\%$ credible interval for the local primordial non-Gaussianity parameter $f_\mr{nl}$ is significantly reduced when the i-trispectrum monopole is added. It further shrinks, even if less substantially, when both bispectrum and i-trispectrum quadrupoles are added to the data-vector.
All the relative improvements on the forecasted parameters constraints are reported in Table \ref{tab:table_constr}.

\begin{figure}[tbp]
\centering 
\includegraphics[width=1.0\textwidth]
{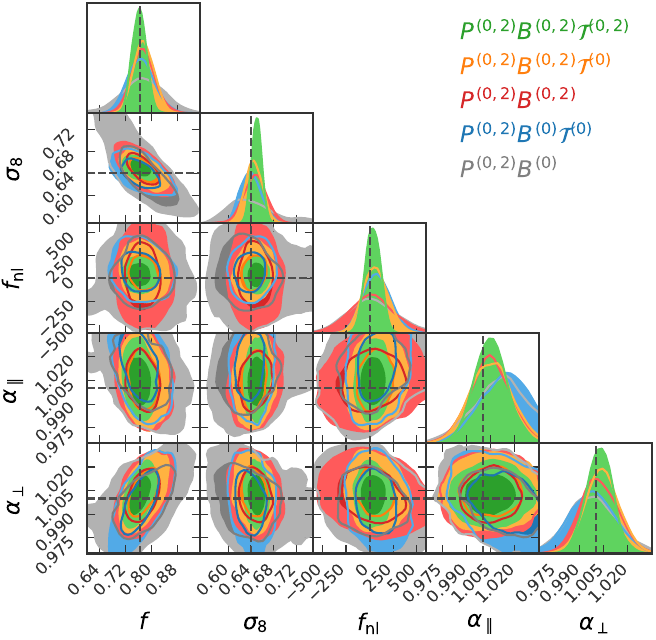}
\caption{
\label{fig:cosmo_p_post}
1-2D 68$\%$ credible regions of the marginalised posterior distributions for the cosmological parameters, obtained using the different statistics combination (for the full plot including also the nuisance parameters see Figure \ref{fig:mcmc_posteriors} in Appendix \ref{app:add_mat}). We progressively added to the joint data-vector (initially power spectrum monopole and quadrupole plus bispectrum monopole): i-trispectrum monopole or bispectrum quadrupoles, both $\mathcal{T}^{(0)}$ and $B^{(2)}$, and finally also the i-trispectrum quadrupoles. The black dashed lines correspond to the true values of the model parameters given by the simulations cosmology presented in Section \ref{sec:analysis_setup}. Recall that the kernel $Z^{(2)}$ was separately calibrated for bispectrum and i-trispectrum using in both cases the monopole and quadrupoles terms together as data-vectors ($B^{(0,2)}$ and $\mathcal{T}^{(0,2)}$), as described in Appendix \ref{sec:app_fit_ker}. This explains the small bias (within the 1D $68\%$ credible regions) with respect to the true simulations parameters present when only the bispectrum or i-trispectrum monopoles are considered.
}
\end{figure}

\subsection{MCMC sampling}
\label{sec:mcmc_sampling}
Using the same data-vector combinations of $P^{(0,2)}$, $B^{(0,2)}$ and $\mathcal{T}^{(0,2)}$, 
we run MCMCs to derive the parameters multi-dimensional posterior distribution. As in the previous section, the covariance matrix is estimated from 8000 simulations and renormalised to simulate a volume of 25 $(h^{-1}\mr{Gpc})^3$.

In Figure \ref{fig:cosmo_p_post} the marginalised 1-2D posterior distributions for the cosmological parameters are displayed, while the improvements on the constraints with respect to the baseline data-vector $\left[P^{(0,2)},B^{(0)}\right]$ are reported in Table \ref{tab:table_constr}. As expected, the improvements computed by the MCMC analysis are slightly smaller than those obtained using the Fisher formalism in Section \ref{sec:snratio_res}. 
Nevertheless, there are no qualitative differences: the different statistics effects, when added to the joint data-vector, follow the same trend. The complementarity in terms of information content between bispectrum and i-trispectrum multipoles is confirmed by the MCMC analysis\footnote{ This was not granted from the Fisher formalism analysis results, where the forecasted marginalised 1$\sigma$ intervals are derived under the approximation of the parameter space being only linearly degenerate.}.

These findings clearly indicate that future analyses targeting primordial non-Gaussianity should prioritise including in their data-vector the isotropic i-trispectrum over the bispectrum quadrupoles. When $\mathcal{T}^{(0)}$ is added to the baseline data-vector, it halves the 1D $68\%$ credible interval for $f_\mr{nl}$, while including the bispectrum quadrupoles reduces it only by $10\%$. In \cite{Gualdi:2020ymf} using a calibrated analytical covariance matrix, we reported a similar effect for the bispectrum quadrupoles, with a maximum improvement of $28.5\%$ when also including all the bispectrum hexadecapole terms.
With the  $\mathcal{T}$-Expreso technique of Appendix \ref{app:fast_tk_model}, including ${\cal T}$ in an MCMC is computationally cheaper than including the bispectrum multipoles, providing a strong motivation for using the i-trispectrum to constrain $f_{\rm nl}$.

The cosmological parameters most frequently constrained by galaxy clustering analyses, $(f,\sigma_8)$ benefit in similar measure from the inclusion of $B^{(0,2)}$ and $\mathcal{T}^{(0,2)}$. Confirming the Fisher forecasts results, we conclude that the anisotropic components of both bispectrum and i-trispectrum are mostly sensitive to the Alcock-Paczy\`{n}ski parameters $(\alpha_\parallel,\alpha_\perp)$.

The promising i-trispectrum quadrupoles signal strength, detected in the (S/N) study of Section \ref{sec:snratio_res} and Figure \ref{fig:signal_to_noise}, translated into a strong reduction ($\sim60\%$ on average) of the 1D $68\%$ credible regions for all the cosmological parameters of interest $(f,\sigma_8,f_\mr{nl},\alpha_\parallel,\alpha_\perp)$. 

In Figure \ref{fig:mcmc_post_fnl_fixed} the results derived from repeating the analysis fixing $f_\mr{nl}=0$ are reported. In Appendix \ref{app:add_mat} Table \ref{tab:table_constr_fnl0} the constraints and relative improvements are compared to the case with $f_\mr{nl}$ free to vary. From both figure and table, it is clear that reducing the degeneracy level in the posterior distribution, by removing the $f_\mr{nl}$ parameter, does not  alter the  results in any significant way.

\renewcommand{\arraystretch}{2.}
\begin{table}[tbp]
\centering
\begin{tabular}{c|cc|cc|cc|cc|cc|}
\cline{2-11}
 & \multicolumn{10}{c|}{MCMC (Fisher Forecasts)} \\
\cline{2-11}
&
 \multicolumn{2}{c|}{$\Delta \theta$}
&
 \multicolumn{8}{c|}{1 - $\left(\Delta\theta/\Delta \theta_{P^{(0,2)}B^{(0)}}\right)$     $\left[\%\right]$}\\
\cline{2-11}
&
\multicolumn{2}{c|}{$P^{(0,2)}B^{(0)}$}  &
\multicolumn{2}{c|}{$P^{(0,2)}B^{(0)}\mathcal{T}^{(0)}$}  & 
\multicolumn{2}{c|}{$P^{(0,2)}B^{(0,2)}$} &
\multicolumn{2}{c|}{$P^{(0,2)}B^{(0,2)}\mathcal{T}^{(0)}$} &
\multicolumn{2}{c|}{$P^{(0,2)}B^{(0,2)}\mathcal{T}^{(0,2)}$} \\
\hline
$f$         & \tf{0.140} & (\tf{0.090}) & 42.8    & (48.8)    & 46.4 & (66.8) & 57.7 & (70.4) & 71.9 & (76.5) \\
$\sigma_8$      & \tf{0.078} & (\tf{0.062}) & 52.3    & (64.1)    & 40.6 & (71.5) & 62.3 & (77.8) & 78.2 & (85.1) \\
$f_\mr{nl}$     & \tf{536}  & (\tf{282})  & 49.1    & (44.4)    & 9.5 & (9.5) & 55.1 & (48.6) & 71.7 & (68.7) \\
$\alpha_\parallel$  & \tf{0.036} & (\tf{0.022}) & 13.9    & (10.6)    & 30.6 & (37.2) & 34.2 & (39)  & 46.8 & (47.7) \\
$\alpha_\perp$    & \tf{0.032} & (\tf{0.018}) & 14.3    & (14.5)    & 29.1 & (35.8) & 33  & (39.2) & 46.3 & (46.3) \\
\hline
\multicolumn{3}{c|}{average improvement}
& 31.3   & (36.5)    & 32.1 & (44.2) & 44.6  & (55) & 61 & (64.9) \\
\hline
\end{tabular}
\caption{\label{tab:table_constr}
Comparison between the 1D $68\%$ credible regions obtained through MCMC sampling and the forecasted ones derived through the Fisher formalism for each statistics combination. In both cases the covariance was re-scaled to simulate a volume of $V_\mr{survey} =25\,(h^{-1}\mr{Gpc})^3$ and corrected to account for the limited number of simulations available through the prescription described in Appendix \ref{app:sel_heav_corr}. This table focuses on the cosmological parameters $(f,\sigma_8,f_\mr{nl},\alpha_\parallel,\alpha_\perp)$, together with their average improvements reported on the last line. In Appendix \ref{app:add_mat}, Table \ref{tab:table_constr_nuisance} the corresponding results for the nuisance parameters $(b_1,b_2,b_3,\sigma_P,\sigma_B,\sigma_\mathcal{T},A_\mr{n})$ are reported.
}
\end{table}

\section{Conclusions}
\label{sec:conclusions}

For the first time, we have presented and assessed the potential of a joint analysis of the anisotropic signal in redshift space of power spectrum, bispectrum and i-trispectrum.

To make this study as relevant as possible for its application to forthcoming data-sets, these statistics were measured from 8000 N-body dark matter simulations at redshift $z~=~0.5$. The catalogues were randomly down-sampled to replicate a realistic object density of $n_\mr{p}~=~5\times10^{-4} (h\,\mr{Mpc}^{-1})^{3}$. The covariance matrix employed in the likelihood evaluation was re-scaled to simulate a survey volume of $V_\mr{survey} =25\,(h^{-1}\mr{Gpc})^3$.

Building on the previous work presented in \cite{Gualdi:2020eag}, we measured and modelled the anisotropic components of both bispectrum and i-trispectrum in redshift space. For the latter, a pipeline ( $\mathcal{T}$-Expreso) was developed to massively speed up the calculation of the  i-trispectrum signal theoretical model, making it possible to include the i-trispectrum in cosmological parameter inference via MCMCs.

A phenomenological model for the bispectrum and trispectrum in the mildly non-linear regime including the shot-noise contributions was presented (see Figures \ref{fig:fit_pbt_stats}, \ref{fig:fit_pbt_ratios} and Appendixes \ref{sec:app_fit_ker}, \ref{sec:app_shot_noise}). This model performs sufficiently well for not biasing the recovery of cosmological parameters from an effective survey volume of $\sim 25 (h^{-1}\rm{Gpc})^3$. However, there is room for improvement: simple modifications or recalibrations of second-order SPT-like kernels fail to reproduce the signal of squeezed or flattened configurations even in the mildly non-linear regime \cite{Lazanu:2015rta,Novell:2021}.

Using only the measurements from the simulations, Figure \ref{fig:signal_to_noise} demonstrates in terms of cumulative signal-to-noise ratio the extra-information contained in both bispectrum and i-trispectrum multipoles. More importantly, it shows their complementarity.

An extensive parameter set was chosen for the analysis. The nuisance parameters $(b_1,b_2,b_3,\sigma_P,\sigma_B,\sigma_\mathcal{T},A_\mr{n})$ are necessary to describe the bias relation, the finger-of-God damping and deviations from a Poisson-like shot-noise. The cosmological parameters $(f,\sigma_8,f_\mr{nl},\alpha_\parallel,\alpha_\perp)$ encode most of the information that clustering analyses aim to probe in a robust way.

\begin{figure}[tbp]
\centering 
\includegraphics[width=1.0\textwidth]
{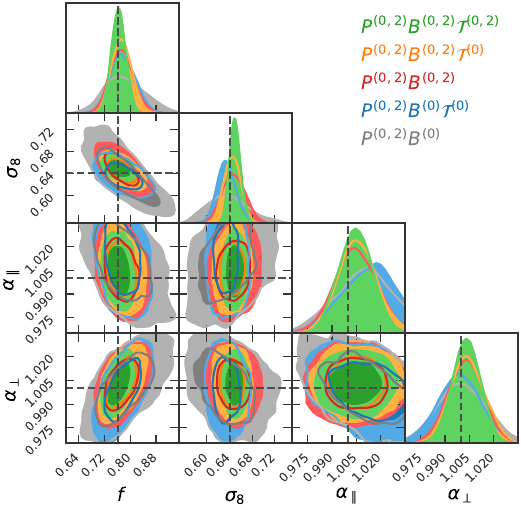}
\caption{\label{fig:mcmc_post_fnl_fixed}
Same as Figure \ref{fig:cosmo_p_post} but for  $f_\mr{nl}\equiv 0$.
The improvements in parameters constraints obtained by adding bispectrum and i-trispectrum multipoles to the data-vector follow the same trend and have similar magnitude as in the full (twelve parameters) case. The quantitative comparison is reported in Table \ref{tab:table_constr_fnl0} in Appendix \ref{app:add_mat}.
}
\end{figure}

The inclusion of the i-trispectrum is particularly effective when considering primordial non-Gaussianity.
 This recovers what we described first in \cite{Gualdi:2020eag}, with the substantial difference of considering a much more degenerate parameter space.
Indeed both i-trispectrum monopole and quadrupoles considerably tighten the forecasted 1D $68\%$
for the primordial non-Gaussianity of the local type amplitude parameter $f_\mr{nl}$, with the caveat that the scale-dependent non-Gaussian halo bias effect has been neglected. It is well known that the halo bias offers an independent route to constrain $f_{\rm nl}$ from the power spectrum of biased tracers.  Nevertheless, should  the proposed clustering analysis be safely extended to larger $k$'s, it may provide constraints on $f_\mr{nl}$ competitive with those obtained from current CMB experiments \cite{Akrami:2019izv},  even without resorting to the scale-dependent halo bias signal. This further highlights the synergy of these different approaches. 

The quadrupoles of both bispectrum and i-trispectrum notably enhance the constraints on the Alcock-Paczy\`{n}ski parameters.
The constraints on the growth rate $f$ and dark matter clustering amplitude normalisation $\sigma_8$ similarly benefit from the addition of bispectrum and i-trispectrum multipoles.

The 1-2D marginalised posterior distributions (sampled through MCMC) for the cosmological parameters are presented in Figure \ref{fig:cosmo_p_post}, while in Appendix \ref{app:add_mat} Figure \ref{fig:mcmc_posteriors} the full parameters set is shown. Table \ref{tab:table_constr} reports the 1D $68\%$ credible regions for the parameters derived with the baseline data-vector $\left[P^{(0,2)}B^{(0)}\right]$ together with the relative improvements obtained by adding the bispectrum quadrupoles and i-trispectrum multipoles.
Figure \ref{fig:mcmc_post_fnl_fixed} and Table \ref{tab:table_constr_fnl0} displays the results derived when fixing $f_\mr{nl}=0$.

When also considering $f_\mr{nl}$ as a free parameter, the sensitivity of the anisotropic bispectrum and i-trispectrum statistics to the different parameters can be summarised as

\begin{itemize}
  \item $\alpha_\parallel$ and $\alpha_\perp$ are better constrained by both anisotropic bispectrum and i-trispectrum ($\sim30\%$ when adding $B^{(2)}$ and $\sim46\%$ when also adding $\mathcal{T}^{(2)}$);
  \item $f$ and $\sigma_8$ constraints are considerably improved by the inclusion of $B^{(2)}$ as well as $\mathcal{T}^{(0,2)}$, reaching 1D $68\%$ credible regions smaller by $\sim 72$\% for $f$ and $\sim 78\%$ for $\sigma_8$ compared to the baseline data-vector;
  \item $f_\mr{nl}$ degeneracy with other parameters is largely alleviated by both the i-trispectrum monopole and quadrupoles, with a globlal improvement in the 1D 68\% credible region of $\sim 72\%$.
\end{itemize}

\noindent The PNG constraints improvement due to the i-trispectrum can be understood by noticing in Equation \ref{eq:png_bktk} the different "response" of $B$ and ${\cal T}$ to $f_\mr{nl}$.
While the signal in the bispectrum is purely primordial, in the i-trispectrum there is a coupling between primordial non-Gaussianity and gravitational evolution, encoded in real space by the kernel $F^{(2)}$. In \cite{Gualdi:2020eag} (figure 4 page 17) we saw that for the bispectrum the relative impact of the primordial term with respect to the gravitational one decreased as $k$ increased, for the i-trispectrum this remained approximately constant over the whole range of configurations (and hence scales).
Compared to the bispectrum primordial term, in redshift space the presence of the kernel $Z^{(2)}$ also implies a more complex dependence of the i-trispectrum on the PNG signal via the skew-quadrilateral orientation with respect to the line of sight. 
In \cite{Gualdi:2020ymf} we found a similar effect, where inclusion of the redshift-space bispectrum anisotropic signal yields an improvement of the $f_\mr{nl}$ constraints, with a maximum error reduction of $\sim 30\%$ when also the hexadecapole terms were added to the data-vector.
In comparison Table \ref{tab:table_constr} shows that only adding the i-trispectrum monopole to the baseline data-vector improves the 1D $68\%$ credible region for $f_\mr{nl}$ by $\sim49\%$. 

It is important to consider that  to achieve an error improvement of $x$\%  (as defined in Table \ref{tab:table_constr}) using only the baseline statistics and without increasing the $k_{\rm max}$, the survey volume would have to be increased by $\sim (1-x/100)^{-2}$.  Hence the improvement in constraints could be quantified with an equivalent volume increase ($V_{\rm eq}$).  
For data-vectors given by different combinations of our statistics, the average improvement on the cosmological parameters reported in Table \ref{tab:table_constr}, with respect to the baseline $\left[P^{(0,2)}B^{(0)}\right]$ case, is: 
\begin{itemize}
  \item 31$\%$ and $V_\mr{eq}\approx 2.1 \,V_\mr{obs}$ for $\left[P^{(0,2)}B^{(0)}\mathcal{T}^{(0)}\right]$ (29$\%$ and $V_\mr{eq}\approx 2\, V_\mr{obs}$ with $f_\mr{nl}$ fixed);
  \item 32$\%$ and $V_\mr{eq}\approx 2.2 \,V_\mr{obs}$ for $\left[P^{(0,2)}B^{(0,2)}\right]$ (36$\%$ and $V_\mr{eq}\approx 2.4 \, V_\mr{obs}$ with $f_\mr{nl}$ fixed);
  \item 45$\%$ and $V_\mr{eq}\approx 3.3 \,V_\mr{obs}$ for $\left[P^{(0,2)}B^{(0,2)}\mathcal{T}^{(0)}\right]$ (44$\%$ and $V_\mr{eq}\approx 3.2\, V_\mr{obs}$ with $f_\mr{nl}$ fixed);
  \item 61$\%$ and $V_\mr{eq}\approx 6.6 \,V_\mr{obs}$ for $\left[P^{(0,2)}B^{(0,2)}\mathcal{T}^{(0,2)}\right]$ (56$\%$ and $V_\mr{eq}\approx 5.2 \, V_\mr{obs}$ with $f_\mr{nl}$ fixed). 
\end{itemize}

\noindent 
We find (see Figure \ref{fig:mcmc_post_fnl_fixed} and Table \ref{tab:table_constr_fnl0} in Appendix \ref{app:add_mat}) that fixing $f_{\rm nl}$ to zero or leaving it as a free parameter does not significantly change the constraints (improvements trend or magnitude) for the other parameters.

The analysis presented here is still idealised in two main aspects: {\it i)} realistic surveys are not perfect boxes, their survey geometry is shaped by selection and window functions, {\it ii)} galaxies inhabit halos, and theirs sampling of the underlying dark matter distribution is not a Poisson process as simulated here. 
Nevertheless we do not expect that the extra layer of complications introduced by these effects would impact our findings in any significant way. Window, selection function and shot-noise affect all statistics and a lot of work in the community and in survey collaborations is going into correctly modelling and accounting for these effects.  These real world effects may affect the signal to noise of a given statistics (especially when shot-noise is high), but should not significantly alter the gain (i.e. relative effect) of including anisotropic bispectrum and i-trispectrum  to the more traditional data-vector.   A much bigger problem would have been if the higher-order statistics here considered were dominated by the shot-noise for an object density similar to the one of current or future surveys. Here we proved that this was not the case.

We conclude with a reflection. 
To date, most of the cosmological constraints from galaxy surveys have been obtained from a single summary statistics, the power spectrum. In this case, to reduce error-bars on cosmology one must increase the survey volume or increase the $k_{\rm max}$. The sky coverage of deep, state-of-the-art galaxy redshfit surveys has increased steadily over the past decade, from $\sim 10^3$ square degrees of 2dF or WiggleZ to $\sim 10^4$ square degrees of SDSS or BOSS \cite{Colless:2001gk,2010MNRAS.401.1429D,2013AJ....145...10D}. 
The current (and next) generation of galaxy redshift surveys cover most\footnote{While the fraction of the celestial sphere covered, $f_{\rm sky}$, is not 1, and closer to 1/4, the fraction of the sky free from foregrounds that can reasonably be used for cosmological galaxy redshift surveys is probably not drastically larger than that. Hence we can say that current surveys already cover the accessible volume in the redshift range they target.} of the observable volume in their targeted redshift range. 
 However when the volume at a given redshift can not be further increased, if mildly-non-linear scales are included, it is important to consider that key cosmological information percolates to higher-order statistics.
 
The analysis of higher-order statistics, especially in redshift space, is much more challenging and, for this reason, not as mainstream. We have shown that higher-order statistics can be modelled reliably and robustly and that cosmological constraints can be improved drastically by considering the anisotropic signal in redshift space of the bispectrum and i-trispectrum: adding these statistics to the baseline data-vector reduces statistical error on the recovered cosmology as much as would increasing quite significantly the survey volume.
The result of this work motivates pursuing the measurement and interpretation of these statistics from observations despite the added challenges.


\acknowledgments 
D.G. and L.V. thank Francisco Villaescusa-Navarro for invaluable help with the extensive use of the Quijote simulations. 
The measurements from the simulations were performed in the TigerCPU cluster at Princeton. L.V. and D.G. acknowledge support of European Unions Horizon 2020 research and innovation programme ERC (BePreSySe, grant agreement 725327). Funding for this work was partially provided by the Spanish MINECO under projects PGC2018-098866-B-I00 FEDER-EU. HGM acknowledges the support from ‘la Caixa’ Foundation (ID100010434) with code LCF/BQ/PI18/11630024.

\appendix

\begin{figure}[t]
\centering 
\includegraphics[width=1.0\textwidth]
{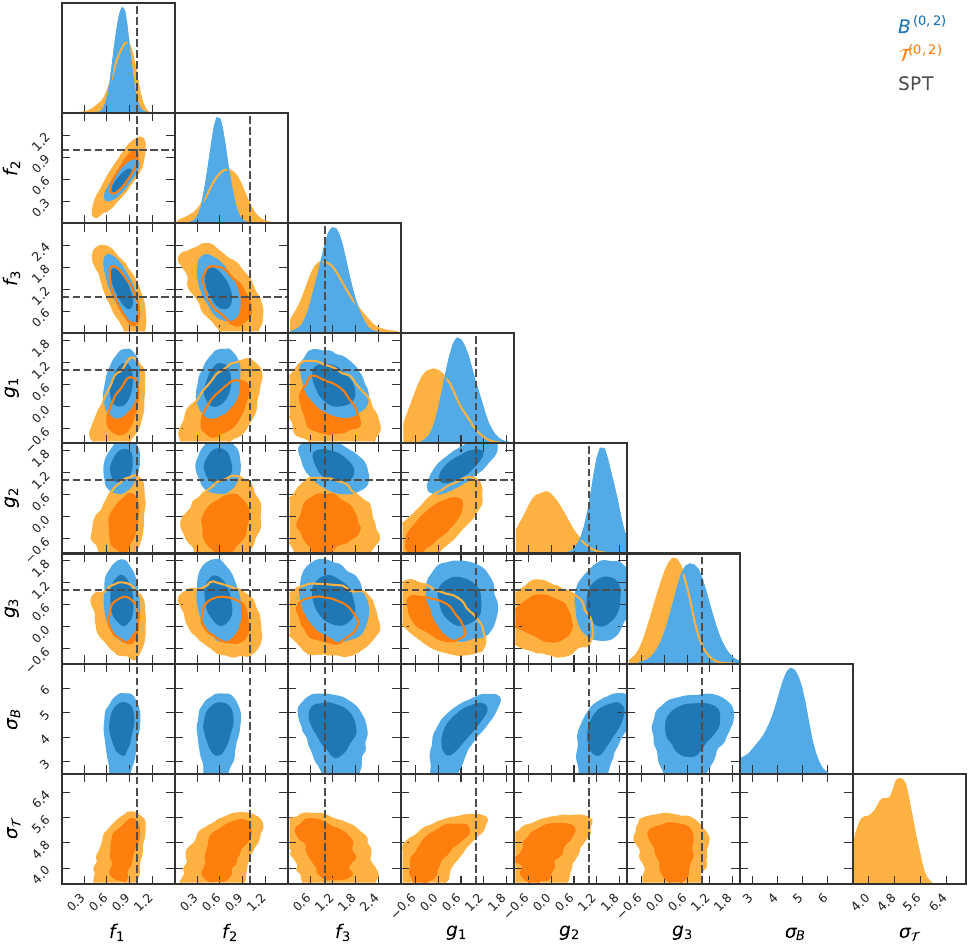}
\caption{\label{fig:posterior_kernel_fit}
1-2D 68$\%$ credible regions of the marginalised posterior distribution of the parameters $(f_1,f_2,f_3,g_1,g_2,g_3,\sigma_B/\sigma_\mathcal{T})$ when fitting both monopole and quadrupoles of either bispectrum or i-trispectrum. The black dashed lines indicate the SPT values (ones). The best-fit parameters for the real space kernel $F^{(2)}$ are compatible between bispectrum and i-trispectrum. On the other hand for the velocity field divergence kernel $G^{(2)}$ the best-fit values in the case of the i-trispectrum are further away from the SPT prediction than for the bispectrum. This should be interpreted as an indicator that further work is necessary to improve the modelling of the i-trispectrum in redshift space. 
}
\end{figure}

\begin{figure}[tbp]
\centering 
\includegraphics[width=1.0\textwidth]
{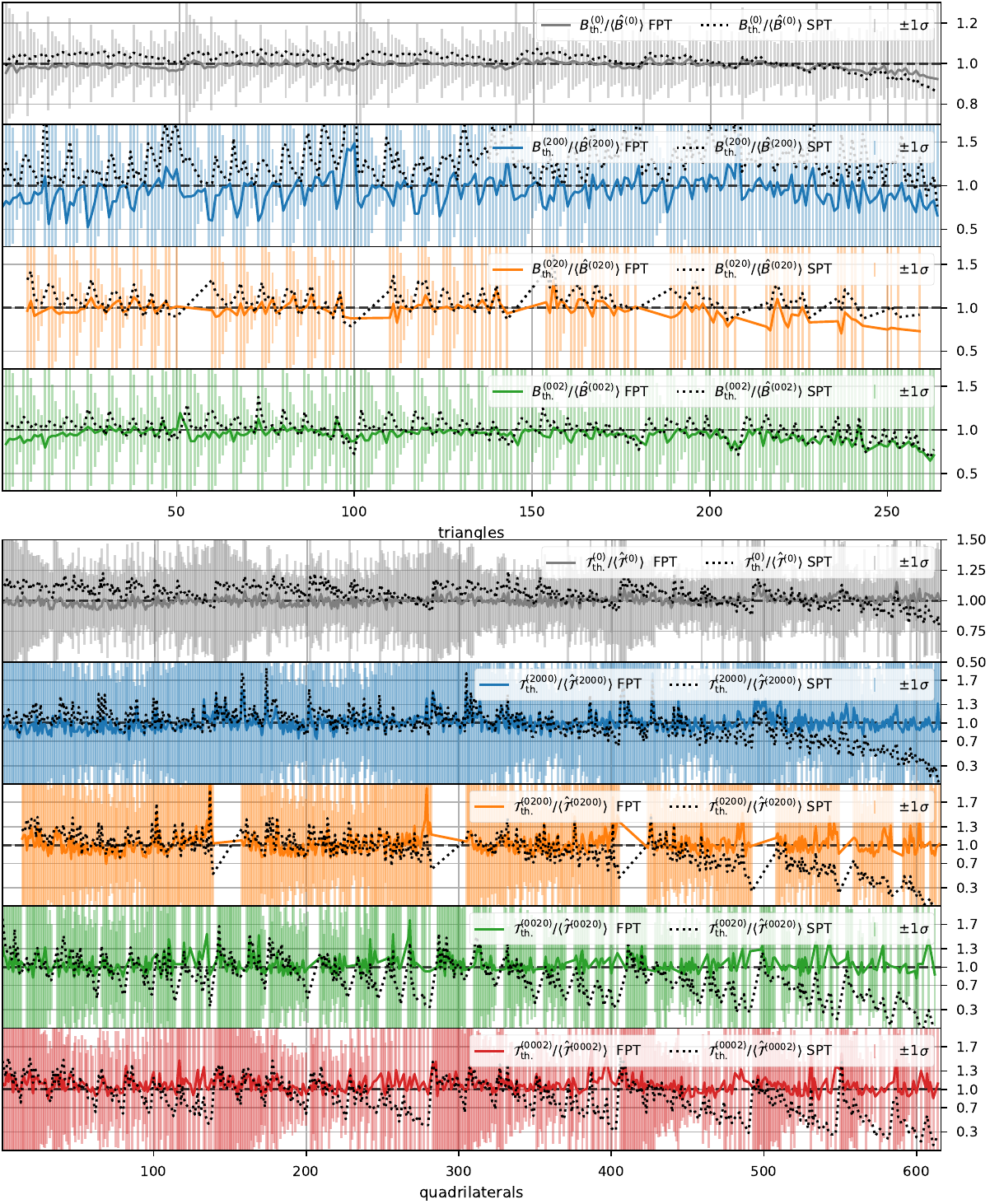}
\caption{\label{fig:kernel_fit}
Comparison between the SPT and FPT best-fit models for the bispectrum multipoles.
}
\end{figure}

\section{Second order SPT kernels fitting}
\label{sec:app_fit_ker}

In order to extend the validity of both bispectrum and i-trispectrum models, we fitted the coefficients of the second-order standard perturbation theory real space kernels of Equation~\ref{eq:2SPT-likekernelsfit} separately for bispectrum and trispectrum. The six parameters $(f_1,f_2,f_3,g_1,g_2,g_3)$ (one set for the bispectrum and another set for the trispectrum) are defined so that when they are all equal to unity, the standard perturbation theory kernels derived in the Einstein-de Sitter Universe approximation are recovered. To fit these parameters we used as data-vector the mean of the 8000 measurements for both the bispectrum and i-trispectrum multipoles, $B^{(0,2)}$ and $\mathcal{T}^{(0,2)}$. The covariance was estimated from the same set of simulations. The theoretical model was computed setting the cosmological parameters to their true values corresponding to the ones used to create the simulations (see Section \ref{sec:simulations} for more details).
In the i-trispectrum case the fitted $F^{(2)}$ and $G^{(2)}$ have been used for both $Z^{(2)}$ and $Z^{(3)}$.

In Figure \ref{fig:posterior_kernel_fit} the 1-2D marginalised posterior distributions for the fitted parameters are shown. The bispectrum best-fit values for the six-parameters are $f_1=0.778,\, f_2=0.540,\, f_3=1.269,\, g_1=0.650,\, g_2=1.458,\, g_3=0.632$ while for the i-trispectrum are $f_1=0.913, \,f_2=0.803,\, f_3=0.902,\, g_1=0.233, \,g_2=0.126, \,g_3=0.127$ .
Without this simple prescription and using the standard perturbation theory kernels, the resulting theoretical model would have provided a poor fit to the data, as can be seen in Figure \ref{fig:kernel_fit} where the best-fit resulting models for SPT (dotted lines, $f_i,g_i=1$ for $i=1,2,3$) and FPT (solid lines) are compared. Especially for the i-trispectrum we can see that for all the multipoles the ratio between theory and average measurement oscillates as a function of the skew-quadrilateral configuration. The same features are present also for the bispectrum but are less prominent. We also found that using SPT kernels even with a lower $k_\mr{max}=0.09\,h\rm{Mpc}^{-1}$ would have resulted in recovering biased best-fit parameters with respect to the true ones corresponding to the simulations cosmology. This does not happen with the suitably calibrated FTP kernels as seen in Section \ref{sec:mcmc_sampling}.

\section{Speeding up the evaluation of the i-trispectrum model: factorisation and grid pre-calculation (\texorpdfstring{$\mf{\mathcal{T}}$}{}-Expreso)}
\label{app:fast_tk_model}

The evaluation of the multi-dimensional integral given in Equation \ref{eq:tk_model} for several quadrilaterals is too computationally expensive and time consuming to be suitable for sampling the parameters posterior with a MCMC.
To bypass this technical bottle-neck we split the integrand of Equation \ref{eq:tk_model} into terms such that the dependence on different parameters combinations can be factorised out.
The various integrals can be suitably pre-computed on a grid, of reduced dimensionality, scanning the remaining parameters.
In the case of the real space kernels fitting step described in Appendix \ref{sec:app_fit_ker}, the i-trispectrum integrand can be decomposed in a total of 35 different integrals depending on different combinations of the parameters $(f_1,f_2,f_3,g_1,g_2,g_3)$. Since we fit the kernels with the model computed at the true cosmology of the simulations, the grid is one-dimensional and the different integrals need to be pre-computed only as a function of $\sigma_\mathcal{T}$.

The same procedure was followed to compute the theoretical model for the i-trispectrum as a function of the bias parameters ($b_1$, $b_2$, $b_3$, $b_\mr{s}$, $b_{\mathcal{G}_2}$, $b_{\mathcal{G}_3}$, $b_{\Gamma_3}$) and the cosmological parameters ($f$,$\sigma_8$,$f_\mathrm{nl}$). The redshift space kernels dependence on different combinations of the bias parameters $b_1$, $b_2$, $b_3$, $b_\mr{s}$, $b_{\mathcal{G}_2}$, $b_{\mathcal{G}_3}$,$b_{\Gamma_3}$ and the growth rate $f$ is ($\sigma_8$ for $B$ and $\mathcal{T}$ acts only as a renormalisation of the matter power spectrum in Equations \ref{eq:bktk_grav}):

\begin{align}
Z^{(1)}\left[\mf{k}\right] &= b_1 + f\,\mu^2\,,
\notag \\
Z^{(2)}\left[\mf{k}_1,\mf{k}_2\right]&=
b_1\,F^{(2)}\left[\mf{k}_1,\mf{k}_2\right]
+ f \mu^2_{12}\,G^{(2)}\left[\mf{k}_1,\mf{k}_2\right]
+ \dfrac{b_2}{2}
+ \dfrac{b_\mr{s}}{2}\,S^{(2)}\left[\mf{k}_1,\mf{k}_2\right]
\notag \\
&
+ \dfrac{b_1f}{2}\left[\mu_1^2 + \mu_2^2 + \mu_1\mu_2\left(\dfrac{k_1}{k_2}+\dfrac{k_2}{k_1}\right)\right]
+f^2 \left[\mu_1^2\mu_2^2 + \dfrac{\mu_1\mu_2}{2}\left(\mu_1^2\dfrac{k_1}{k_2}+\mu_2^2\dfrac{k_2}{k_1}\right)\right]\,,
\notag\\
\notag
\end{align}
\begin{align}
Z^{(3)}\left[\mf{k}_1,\mf{k}_2,\mf{k}_3\right]&=
b_1\,F^{(3)}\left[\mf{k}_1,\mf{k}_2,\mf{k}_3\right]
+ f \mu^2_{123}\,G^{(3)}\left[\mf{k}_1,\mf{k}_2,\mf{k}_3\right]
\notag \\
&+ b_1f\,\Bigg\{
F^{(2)}\left[\mf{k}_2,\mf{k}_3\right]\left[\mu_1^2 + \mu_1\mu_{23}\dfrac{k_{23}}{k_1}\right] + 
G^{(2)}\left[\mf{k}_2,\mf{k}_3\right]\left[\mu_{23}^2 + \mu_1\mu_{23}\dfrac{k_1}{k_{23}}\right]
\Bigg\}
\notag \\
&+
f^2\,G^{(2)}\left[\mf{k}_2,\mf{k}_3\right]
\left(2\mu_1^2\mu_{23}^2+ \mu_1\mu_{23}^3\dfrac{k_{23}}{k_1} + \mu_1^3\mu_{23}\dfrac{k_1}{k_{23}}\right)
\notag \\
&+\left(2\,b_2+4\,b_\mr{s}\,S_2\,\left[\mf{k}_2,\mf{k}_3\right]\right)\,F^{(2)}\left[\mf{k}_2,\mf{k}_3\right]
\notag \\
&+
\dfrac{b_2 f}{2}\left(\mu_1^2 + \mu_1\mu_{23}\dfrac{k_{23}}{k_1}\right)
+b_\mr{s}f\,S^{(2)}\,\left[\mf{k}_2,\mf{k}_3\right]
\left(\mu_1^2 + \mu_1\mu_{23}\dfrac{k_{23}}{k_1}\right)
\notag \\
&+
b_1f^2\,\left[
\mu_2^2\mu_3^2 + \mu_1\mu_2^2\mu_3\dfrac{k_3}{k_1} + \mu_1\mu_3^2\mu_2\dfrac{k_2}{k_1}
+
\dfrac{1}{2}\left(\mu_1\mu_2^3\dfrac{k_2}{k_1}+ \mu_1\mu_3^3\dfrac{k_3}{k_1}+
\mu_2\mu_3\mu_1^2\dfrac{k_1^2}{k_2k_3}
\right)
\right]
\notag \\
&+f^3\,\left[
\mu_1^2\mu_2^2\mu_3^2 
+ \dfrac{3}{2}\left(
\mu_1\mu_2^2\mu_3^3\dfrac{k_3}{k_1} + \mu_1\mu_2^3\mu_3^2\dfrac{k_2}{k_1}
\right)
+\dfrac{1}{2}\mu_1^4\mu_2\mu_3\dfrac{k_1^2}{k_2k_3}
\right]
\notag \\
&+b_3
+b_{\mathcal{G}_2}\,\mathcal{G}_2\left[\mf{k}_2,\mf{k}_3\right]
+b_{\mathcal{G}_3}\,\mathcal{G}_3\left[\mf{k}_1,\mf{k}_2,\mf{k}_3\right]
+b_{\Gamma_3}\,\Gamma_3\left[\mf{k}_1,\mf{k}_2,\mf{k}_3\right]\,.
\label{eq:zkernels}
\end{align}

\noindent $F^{(3)}$ and $G^{(3)}$ are the third order SPT kernels for the over-density and velocity divergence fields, while $S^{(2)}\left[\mf{k}_a,\mf{k}_b\right]=(\mf{k}_a\mf{k}_b)^2/(k_ak_b)^2 - 1/3$. The third order bias parameters ($b_{\mathcal{G}_2}$, $b_{\mathcal{G}_3}$, $b_{\Gamma_3}$) and related kernels are given in \cite{Abidi:2018eyd}, assuming local bias in Lagrangian space these parameters can be expressed in terms of $b_1$, $b_2$ and $b_3$ as for $b_\mr{s}=-\dfrac{4}{7}\left(b_1-1\right)$ \cite{Chan:2012jj,Baldauf:2012hs}.

Considering the two terms making up the gravitational collapse component of the trispectrum, one proportional to $\propto Z^{(1)}Z^{(1)}Z^{(2)}Z^{(2)}$ and the other to $\propto Z^{(1)}Z^{(1)}Z^{(1)}Z^{(3)}$, we find that they can be decomposed into 56 terms, each of them proportional to a different combination of the bias parameters and $f$. For the primordial non-Gaussian signal contained in the i-trispectrum there are a total of 18 different integrals that can be pre-computed in the same way.

As a short notation each term will be labelled using only the specific parameters combination in front of it, for example $b_1f$. To differentiate between terms belonging to different kernels we will write, $\wh{b_1f}$ for $Z_2$'s terms and $\ol{b_1f}$ for $Z_3$. Then a decomposition for the trispectrum gravitational component reads (analogously to what done in \cite{Sugiyama:2020uil} for the bispectrum):
\newpage
\begin{align}
   T^\mr{s}\left(\mf{k}_1,\mf{k}_2,\mf{k}_3,\mf{k}_4\right) &=
 b_1^2\wh{b_1}^2 + b_1^3\ol{b_1} 
 \notag \\ &+
 2b_1^2\wh{b_1}\wh{f} + 2b_1f\wh{b_1}^2 +b_1^3\ol{f} + 3b_1^2f\ol{b_1}
 \notag \\ &+
 2b_1^2\wh{b_1}\wh{f^2}+ 2b_1^2\wh{f}\wh{b_1f} + 4 b_1f\wh{b_1}\wh{b_1f}
 +b_1^3\ol{f^2} + 3b_1^2f\ol{b_1f}
 \notag \\ &+
 2f^2\wh{b_1}\wh{f} + 2b_1f\wh{f}^2 + f^3\ol{b_1} + 3b_1f^2\ol{f}
 \notag \\ &+
 f^2\wh{f}^2 + f^3\ol{f}
 \notag \\ &+
 2f^2\wh{f}\wh{f^2} + f^3\ol{f^2}
 \notag \\ &+
 2b_1^2\wh{b_1f}\wh{f^2} + 2b_1f\left(\wh{b_1f}\right)^2 +
 b_1^3\ol{f^3} + 3b_1^2f\ol{b_1f^2}
 \notag \\ &+
 b_1^2\wh{f^2}^2 + f^2\left(\wh{b_1f}\right)^2 + 4b_1f\wh{b_1f}\wh{f^2}
 +3b_1^2f\ol{f^3} + 3b_1f^2\ol{b_1f^2}
 \notag \\ &+
 2f^2\wh{b_1}\wh{f^2} + 2f^2\wh{f}\wh{b_1f} + 4b_1f\wh{f}\wh{f^2}+
 f^3\ol{b_1f} + 3b_1f^2\ol{f^2}
 \notag \\ &+
 b_1^2\wh{f}^2 + f^2\wh{b_1}^2 + 4b_1f\wh{b_1}\wh{f}
 +3b_1^2f\ol{f} + 3b_1f^2\ol{b_1}
 \notag \\ &+
 2b_1^2\wh{f}\wh{f^2} + 2f^2\wh{b_1}\wh{b_1f} + 4b_1f\wh{b_1}\wh{f^2} + 4b_1f\wh{f}\wh{b_1f}
 +3b_1^2f\ol{f^2} + 3b_1f^2\ol{b_1f}
 \notag \\ &+
2f^2\wh{b_2}\wh{b_1f} + 4b_1f\wh{b_2}\wh{f^2} + 3b_1f^2\ol{b_2f}
\quad + \quad
2f^2\wh{b_\mr{s}}\wh{b_1f} + 4b_1f\wh{b_\mr{s}}\wh{f^2} + 3b_1f^2\ol{b_\mr{s}f}
\notag \\ &+
2f^2\wh{b_1}\wh{b_2} + 4b_1f\wh{f}\wh{b_2} + 3b_1f^2\ol{b_2}
\quad + \quad
2f^2\wh{b_1}\wh{b_\mr{s}} + 4b_1f\wh{f}\wh{b_\mr{s}} + 3b_1f^2\ol{b_\mr{s}}
\notag \\ &+
2b_1^2\wh{f}\wh{b_2} + 4 b_1f\wh{b_1}\wh{b_2} + 3b_1^2f\ol{b_2}
\quad + \quad
2b_1^2\wh{f}\wh{b_\mr{s}} + 4 b_1f\wh{b_1}\wh{b_\mr{s}} + 3b_1^2f\ol{b_\mr{s}}
\notag \\ &+
2b_1^2\wh{b_1}\wh{b_1f} + b_1^3\ol{b_1f}
\notag \\ &+
2b_1^2\wh{b_2}\wh{f^2}+ 4b_1f\wh{b_2}\wh{b_1f} + 3b_1^2f\ol{b_2f}
\quad + \quad
2b_1^2\wh{b_\mr{s}}\wh{f^2}+ 4b_1f\wh{b_\mr{s}}\wh{b_1f} + 3b_1^2f\ol{b_\mr{s}f}
\notag \\ &+
2b_1^2\wh{b_2}\wh{b_1f} + b_1^3\ol{b_2f}
\quad + \quad
2b_1^2\wh{b_\mr{s}}\wh{b_1f} + b_1^3\ol{b_\mr{s}f}
\notag \\ &+
b_1^2\left(\wh{b_1f}\right)^2 + b_1^3\ol{b_1f^2}
\notag \\ &+
2f^2\wh{b_2}\wh{f^2} + f^3\ol{b_2f}
\quad + \quad
2f^2\wh{b_\mr{s}}\wh{f^2} + f^3\ol{b_\mr{s}f}
\notag \\ &+
2f^2\wh{f}\wh{b_2} +f^3\ol{b_2}
\quad + \quad
2f^2\wh{f}\wh{b_\mr{s}} +f^3\ol{b_\mr{s}}
\notag \\ &+
2f^2\wh{b_1f}\wh{f^2}+2b_1f\wh{f^2}^2+f^3\ol{b_1f^2} +3 b_1f^2\ol{f^3}
\notag \\ &+
2b_1^2\wh{b_1}\wh{b_2} + b_1^3\ol{b_2}
\quad + \quad
2b_1^2\wh{b_1}\wh{b_\mr{s}} + b_1^3\ol{b_\mr{s}}
\notag \\ &+
f^2\wh{f^2}^2 +f^3\ol{f^3}
\notag \\ &+
b_1^2\wh{b_2}^2
\quad + \quad
2b_1^2\wh{b_2}\wh{b_\mr{s}}
\quad + \quad
b_1^2\wh{b_\mr{s}}^2
\notag \\ &+
f^2\wh{b_2}^2
\quad + \quad
2f^2\wh{b_2}\wh{b_\mr{s}}
\quad + \quad
f^2\wh{b_\mr{s}}^2
\notag \\ &+
2b_1f\wh{b_2}^2
\quad + \quad
4b_1f\wh{b_2}\wh{b_\mr{s}}
\quad + \quad
2b_1f\wh{b_\mr{s}}^2
\notag \\ &+
b_1^3\left(\ol{b_{\mathcal{G}_2}} \quad + \quad\ol{b_{\mathcal{G}_3}} \quad + \quad \ol{b_{\Gamma_3}}\right)
\notag \\ &+
f^3\left(\ol{b_{\mathcal{G}_2}} \quad + \quad\ol{b_{\mathcal{G}_3}} \quad + \quad \ol{b_{\Gamma_3}}\right)
\notag \\ &+
3b_1^2f\left(\ol{b_{\mathcal{G}_2}} \quad + \quad\ol{b_{\mathcal{G}_3}} \quad + \quad \ol{b_{\Gamma_3}}\right)
\notag \\ &+
3b_1f^2\left(\ol{b_{\mathcal{G}_2}} \quad + \quad\ol{b_{\mathcal{G}_3}} \quad + \quad \ol{b_{\Gamma_3}}\right)\,
\notag \\ &+
b_3\left(\ol{b_{\mathcal{G}_2}} \quad + \quad\ol{b_{\mathcal{G}_3}} \quad + \quad \ol{b_{\Gamma_3}}\right)\,,
\end{align}

\noindent where for the sake of shortness some similar terms have been written on the same line with simply a larger spacing. The primordial trispectrum term decomposition reads:

\begin{align}
  T^\mr{s}_\mr{PNG}\left(\mf{k}_1,\mf{k}_2,\mf{k}_3,\mf{k}_4\right) &=
  b_1^3\wh{b_1}
  \;+\; b_1^3\wh{f}
  \;+\; b_1^3\wh{b_2}
  \;+\; b_1^3\wh{b_\mr{s}}
  \;+\; b_1^3\wh{b_1f}
  \;+\; b_1^3\wh{f^2}
  \notag \\ &+
  f^3\wh{b_1} 
  \;+\; f^3\wh{f}
  \;+\; f^3\wh{b_2}
  \;+\; f^3\wh{b_\mr{s}}
  \;+\; f^3\wh{b_1f}
  \;+\; f^3\wh{f^2}
  \notag \\ &+
  3b_1^2f\wh{b_1}
  \;+\; 3b_1^2f\wh{f} 
  \;+\; 3b_1^2f\wh{b_2}
  \;+\; 3b_1^2f\wh{b_\mr{s}}
  \;+\; 3b_1^2f\wh{b_1f}
  \;+\; 3b_1^2f\wh{f^2}
  \notag \\ &+
  3b_1f^2\wh{b_1}
  \;+\; 3b_1f^2\wh{f}
  \;+\; 3b_1f^2\wh{b_2}
  \;+\; 3b_1f^2\wh{b_\mr{s}}
  \;+\; 3b_1f^2\wh{b_1f}
  \;+\; 3b_1f^2\wh{f^2}\,.
  \notag \\
\end{align}

The grid of pre-computed integrals is three-dimensional since both Alcock-Paczy\`{n}ski parameters and trispectrum FoG parameter $\sigma_\mathcal{T}$ cannot be factorised as for the bias parameters, being coupled with the angular dependence of the $k$-vectors with the line of sight. For $\alpha_\parallel$ and $\alpha_\perp$ the grid spans the interval $\left[0.96,1.04\right]$ while for $\sigma_\mathcal{T}$ the range is $\left[3.5,6.5\right]$. Of course, for a practical application to real data where the best fit parameters are not known a priori, these ranges and grid sampling would need to be extended. In the specific case of this analysis, having to compute more than 70 integrals for the monopole and quadrupoles of each of the 700 quadrilaterals at every node of the grid, the range and number of grid points for each parameter was chosen as small as possible to save computational time. This corresponded to computing all the terms for $9^3=729$ nodes.

\section{Shot-noise terms}
\label{sec:app_shot_noise}

It is well known that  the  power spectrum, bispectrum and i-trispectrum of discrete distributions have  additional "shot noise" contributions, which  as often refereed to as $P_{\rm SN}$, $B_{\rm SN}$ and ${\cal T}_{\rm SN}$ respectively. For Poisson shot noise these terms' expressions (e.g., \cite{Verde:2001pf}) are proportional to powers of $1/n_{\rm p}$ where $n_{\rm p}$ denotes the number density of tracers, or, in this case particles. 

To account for the shot-noise contribution in all the measured statistics, we followed two separate approaches both based on corrections given in the literature e.g., \cite{Verde:2001pf}. For both power spectrum and bispectrum the shot-noise was directly measured from each individual simulation and subsequently subtracted from the measured $n$-point correlator in Fourier space:
\begin{align}
\label{eq:pb_shotnoise}
  \hat{P}^{(0)}(k) &= \langle\delta^{(0)}(k)\delta^{(0)}(k)\rangle - \dfrac{1}{n_\mr{p}}\,,
  \notag \\
  \hat{P}^{(2)}(k) &= \langle\delta^{(2)}(k)\delta^{(0)}(k)\rangle\,;
  \notag \\
  \hat{B}^{(0)}(k_1,k_2,k_3) &= \langle\delta^{(0)}(k_1)\delta^{(0)}(k_2)\delta^{(0)}(k_3)\rangle - \left(\hat{P}^{(0)}(k_1)+\hat{P}^{(0)}(k_2)+\hat{P}^{(0)}(k_3)\right)\dfrac{1}{n_\mr{p}}+\dfrac{1}{n_\mr{p}^2}\,
  \notag \\
  \hat{B}^{(200)}(k_1,k_2,k_3) &= \langle\delta^{(2)}(k_1)\delta^{(0)}(k_2)\delta^{(0)}(k_3)\rangle - \hat{P}^{(2)}(k_1)\dfrac{1}{n_\mr{p}}\,.
\end{align}
For the i-trispectrum, the shot-noise correction contains terms proportional to the bispectrum which in turns depends on the possible values of the skew-quadrilateral diagonal $D$. Instead of using the average value of $D$ as done in \cite{Gualdi:2020eag}, we opted for the more precise choice of using an analytical expression, integrating over all the possible values of $D$:
\begin{align}
\label{eq:tk_shotnoise}
\hat{\mathcal{T}}^{(0)}(k_1,k_2,k_3,k_4) &= \langle\delta^{(0)}(k_1)\delta^{(0)}(k_2)\delta^{(0)}(k_3)\delta^{(0)}(k_4)\rangle 
\notag \\ &- 
\dfrac{1}{3}\sum_{\substack{k_1,k_2,k_3,k_4 \\ k_1,k_3,k_2,k_4 \\ k_1,k_2,k_4,k_3}}
\dfrac{1}{16\pi^2 \Delta D}
\int^{D_\mr{max}}_{D_\mr{min}}dD\int^{+1}_{-1}\,d\mu_D\int^{2\pi}_0\,d\phi_{12}\int^{2\pi}_0\,d\psi
\notag \\
&\times\Bigg\{
\dfrac{1}{n_\mr{p}}\left(
B(\mf{k}_1,\mf{k}_2,\mf{k}_3 + \mf{k}_4) +\,5\,\mr{p.}\right)\,
\notag \\
&+
\dfrac{1}{n_\mr{p}^2}
\left(
P(\mf{k}_1)+\,3\,\mr{p.} \,+
P(\mf{k}_1+\mf{k}_2)+\,5\,\mr{p.}
\right)
+ \dfrac{1}{n_\mr{p}^3}
\Bigg\}
\,
\notag \\
\notag
\end{align}
\begin{align}
\hat{\mathcal{T}}^{(2000)}(k_1,k_2,k_3,k_4) &= \langle\delta^{(2)}(k_1)\delta^{(0)}(k_2)\delta^{(0)}(k_3)\delta^{(0)}(k_4)\rangle
\notag \\&- 
\dfrac{1}{3}\sum_{\substack{k_1,k_2,k_3,k_4 \\ k_1,k_3,k_2,k_4 \\ k_1,k_2,k_4,k_3}}
\dfrac{1}{16\pi^2 \Delta D}
\int^{D_\mr{max}}_{D_\mr{min}}dD\int^{+1}_{-1}\,d\mu_D\int^{2\pi}_0\,d\phi_{12}\int^{2\pi}_0\,d\psi \, \mathcal{L}_2(\mu_1)
\notag \\ 
&\times
\Bigg\{\dfrac{1}{n_\mr{p}}\left(
B(\mf{k}_1,\mf{k}_2,\mf{k}_3 + \mf{k}_4) +\,5\,\mr{p.}\right)\,
\notag \\
&+
\dfrac{1}{n_\mr{p}^2}
\left(
P(\mf{k}_1)+\,3\,\mr{p.} \,+
P(\mf{k}_1+\mf{k}_2)+\,5\,\mr{p.}
\right)
+ \dfrac{1}{n_\mr{p}^3}
\Bigg\}
\,.
\end{align}
\noindent Notice that, in real data analysis situation, a more flexible approach would consist in using an analytical prescription for the shot-noise correction of all the statistics, recomputing it at each likelihood evaluation as already done for the theoretical model of each statistics.

\begin{figure}[tbp]
\centering 
\includegraphics[width=1.0\textwidth]
{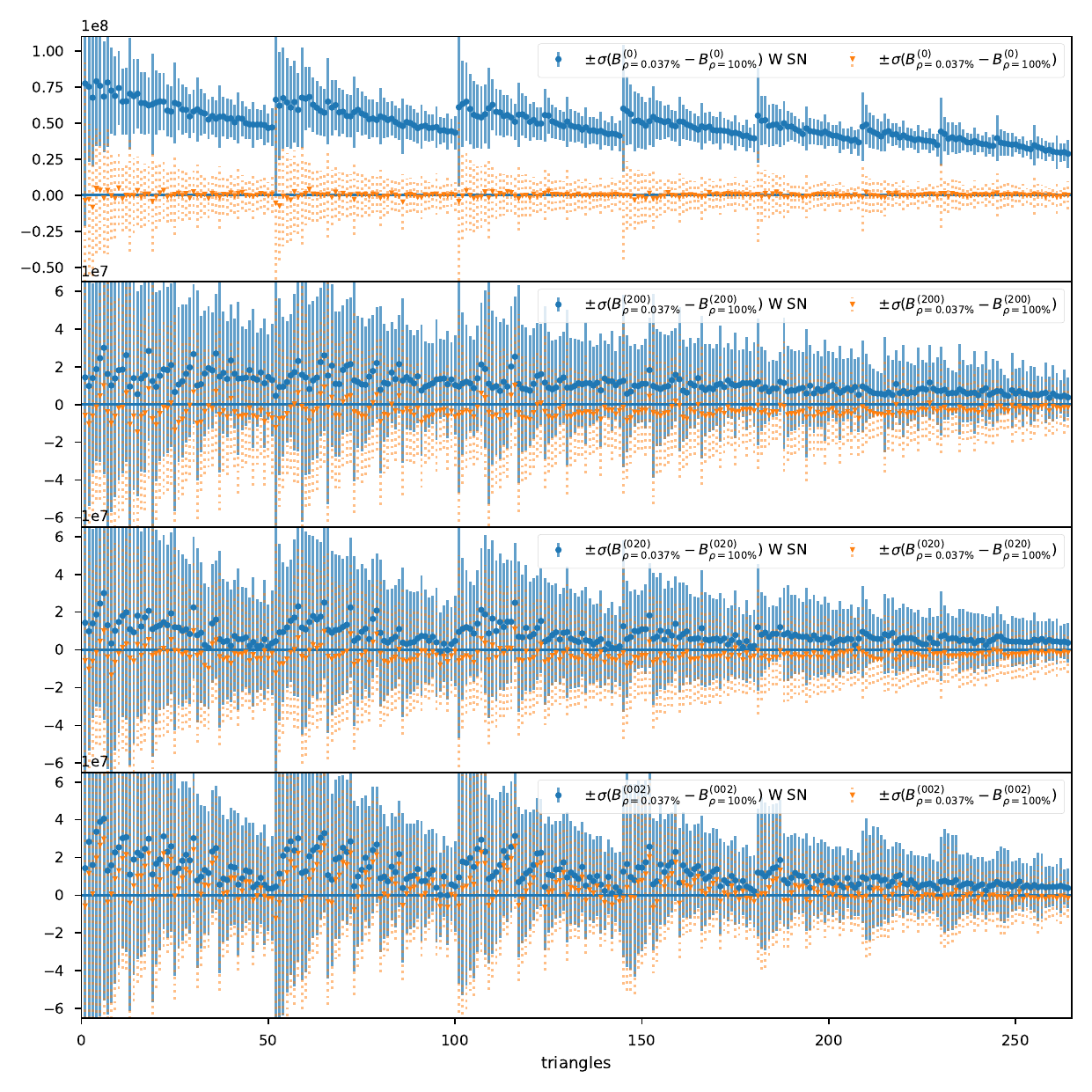}
\caption{\label{fig:sn_bispectrum} 
Difference for the bispectrum multipoles between measurements on the full density and randomly undersampled simulations ($0.037\%$ of the original number of particles). Two cases are shown: before (blue) and after (orange) shot-noise term subtraction. The blue dots and orange triangles represent the mean of the differences for the 8000 simulations while the error-bars correspond to the standard deviation of the difference for each quadrilateral. Especially for the bispectrum monopole subtracting the shot-noise has a not-negligible impact in correctly measuring the statistic.
}
\end{figure}

\begin{figure}[tbp]
\centering 
\includegraphics[width=1.0\textwidth]
{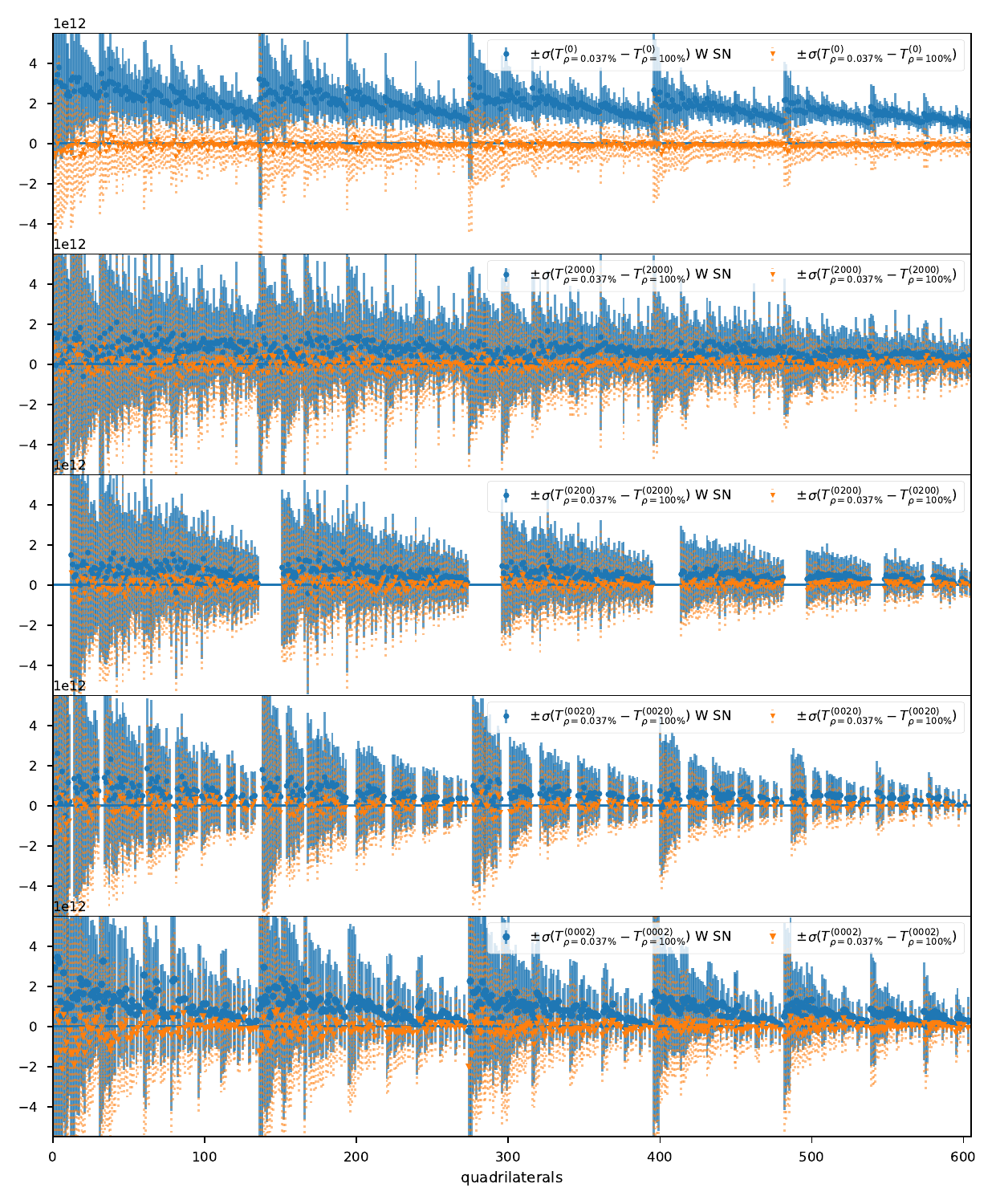}
\caption{\label{fig:sn_trispectrum}
Same as Figure \ref{fig:sn_bispectrum} but for the i-trispectrum case. Also in this case subtracting the shot-noise term as described in equation \ref{eq:tk_shotnoise} allows to recover from the undersampled catalogues compatible measurements of the i-trispectrum with the ones obtained using the full density catalogues.
}
\end{figure}

To check the accuracy of the shot-noise subtraction from the measurements, we measure the same statistics from the simulations at full-density. Using all the dark matter particles available, the average density is $\sim 1000$ larger and hence the shot-noise correction is negligible with respect to the signal. 
In Figures \ref{fig:sn_bispectrum}, for the bispectrum, and \ref{fig:sn_trispectrum}, for the i-trispectrum,  the difference between the statistics  measured on the down-sampled and full-density simulations, before shot-noise subtraction is shown in blue. In orange the same differences are displayed after the shot-noise removal from the measurements. 
these figures show that the adopted shot-noise prescription is satisfactory in reproducing the signal as measured from the full-density simulations. 

In the modelling for the different terms, to account for deviations from Poissonian shot-noise we included the parameter $A_\mr{n}$ in the following way:

\begin{align}
\label{eq:an_param}
    P &\longrightarrow P + \dfrac{A_\mr{n}}{\alpha_\parallel\alpha_\perp^2} \times P_\mr{SN} 
    \notag \\
    B &\longrightarrow B + \dfrac{A_\mr{n}}{\alpha_\parallel^2\alpha_\perp^4} \times B_\mr{SN}
    \notag \\
    \mathcal{T} &\longrightarrow \mathcal{T} + \dfrac{A_\mr{n}}{\alpha_\parallel^3\alpha_\perp^6} \times \mathcal{T}_\mr{SN}\,,
\end{align}

\noindent where the expressions for  $P_\mr{SN}$, $B_\mr{SN}$ and $\mathcal{T}_\mr{SN}$  can be obtained by looking at the  terms on the right hand sides of Equations \ref{eq:pb_shotnoise} and \ref{eq:tk_shotnoise} proportional to powers of $n_\mr{p}$.

\section{Sellentin-Heavens vs. Hartlap correction}
\label{app:sel_heav_corr}
It is well known that if the covariance for a data-vector of size $p$ is estimated by a set of $n$ simulations (or realisations) where $n>p$, the estimated covariance will be approximated, the approximation being increasingly better in the regime where $n\gg p$. Several corrections have been proposed to account for this, the most widely used being the Hartlap correction \cite{Hartlap:2006kj}. Reference~\cite{Sellentin:2015waz} on the other hand is the definitive treatment on this issue, demonstrating that the full form of the likelihood need to be corrected (SH). Here we briefly review the two approaches and show the correspondence between the two corrections to the level of the curvature of the likelihood at its maximum. In Section \ref{sec:snratio_res} when implementing the (S/N) estimate and the Fisher matrix forecasts we use the Hartlap correction, while in Section \ref{sec:mcmc_sampling} when performing MCMCs we correct the likelihood by implementing SH. This appendix motivates this choice. 

According to the Hartlap prescription, if the numerically-estimated covariance is $\Sigma$ then its inverse needs to be corrected by the factor $a_H=(n-p-2)/(n-1)$. In practice the correction is applied to the Likelihood or the $\chi^2$ as: 
\begin{equation}
\ln L={\cal C}-\frac{a_H}{2}\chi^2\,,
\end{equation}
 where ${\cal C}$ denotes a constant. For $n\gg p$, $a_H\longrightarrow 1$.

Note that since the correction is a constant factor that does not depend on the theory model, the parameters values etc. the Hartlap correction just re-scales the covariance by $1/a_H$, and this just re-scales the errors by $1/\sqrt{a_H}$. For example when doing Fisher matrix-based error forecasts with numerically estimated covariances, by applying the Hartlap correction, the errors will be re-scaled by  $1/\sqrt{a_H}$, which for $n\gg1$ and $n\gg p$ gives 

\begin{equation}
a_H \sim 1-\frac{p}{n} \,\,\,\,{\rm and}\,\,\,{\rm thus}\,\,\, \frac{1}{\sqrt{a_H}}\sim 1+\frac{1}{2}\frac{p}{n}. 
\end{equation}
Hence errors become bigger compared to the naive estimate. 

 In the SH approach ~\cite{Sellentin:2015waz}, the correction is to the likelihood and is given by\footnote{For a clear and succinct summary see https://www.imperial.ac.uk/people/a.heavens/research.html under "The Hartlap Correction"}:
\begin{equation}
\label{eq:like_correction_sh}
\ln L={\cal C}'-\frac{n}{2}\ln\left[1+\frac{\chi^2}{(n-1)}\right]\,,
\end{equation}
where ${\cal C}'$ is a constant. Note that it seems that the explicit dependence on $p$ has disappeared (but read on).

One could be tempted to apply this correction to Fisher-based forecasts considering that the Fisher matrix, $F_{ij}$, is given by the second derivative of the log-likelihood evaluated at the maximum.

\begin{equation}
F_{ij}=\frac{n}{2}\left.\frac{\partial^2_{ij}\chi^2}{(n-1)(n-1+\chi^2)}\right|_{max L}\,.
\end{equation}

So if we can say that at the maximum likelihood $\chi^2=p$ then for Fisher the SH correction becomes:
\begin{equation}
\label{eq:sh_factor_approx}
a_{SH}=\frac{n}{(n-1)(1+p/(n-1))}= \frac{n}{(n-1+p)}\longrightarrow_{n\gg1}\frac{n}{p+n}\sim \frac{1}{1+p/n}\,,
\end{equation}
and the explicit dependence on $p$ has reappeared.

In analogy to the Hartlap case, for $p\gg n$ we obtain that:
\begin{equation}
\frac{1}{\sqrt{a_{SH}}}\sim 1+\frac{1}{2}\frac{p}{n}\,.
\end{equation}

In general:
\begin{equation}
\frac{a_{SH}}{a_H}=\left(1-\frac{p+2}{n}\right)\left(1+\frac{p}{n-1}\right)\,,
\label{eq:ratio_1}
\end{equation}
which is always $\le 1$.

In the regime where  $n\gg1$ and $p \gg 2$, because this is the challenging regime, where not necessarily $n\gg p$ but still $p/n<1$, we can write:
\begin{equation}
\frac{a_{SH}}{a_H}\sim\left (1-\frac{p}{n}\right)\left(1+\frac{p}{n}\right)=\left(1-\left(\frac{p}{n}\right)^2\right)\,.
\label{}
\end{equation}
In rescaling the Fisher-estimated errors the quantity that matters is 
\begin{equation}
\frac{\sigma_{SH}}{\sigma_{H}}=\sqrt{\frac{a_{H}}{a_{SH}}}\sim 1+\frac{1}{2}\left(\frac{p}{n}\right)^2\,,
\end{equation}
if $p/n$ sufficiently smaller than 1.
This simply says that the curvature of the SH likelihood is larger than for a Gaussian likelihood, it however does not mean that the 1,2 or 3 $\sigma$ errors are smaller: the SH distribution is more peaked at the core and has broader wings  than the Gaussian Hartlap approximation. \cite{Sellentin:2016psv} shows that the 1-sigma confidence level of Hartlap and the full SH likelihood are actually quite similar. However the (correct) SH likelihood yields larger errors at the 2 $\sigma$ level and beyond. 

For this reason we use the Hartlap correction for the Fisher and signal-to-noise calculations but we use the correct SH likelihood for the MCMC runs. 

\begin{figure}[tbp]
\centering 
\includegraphics[width=1.\textwidth]
{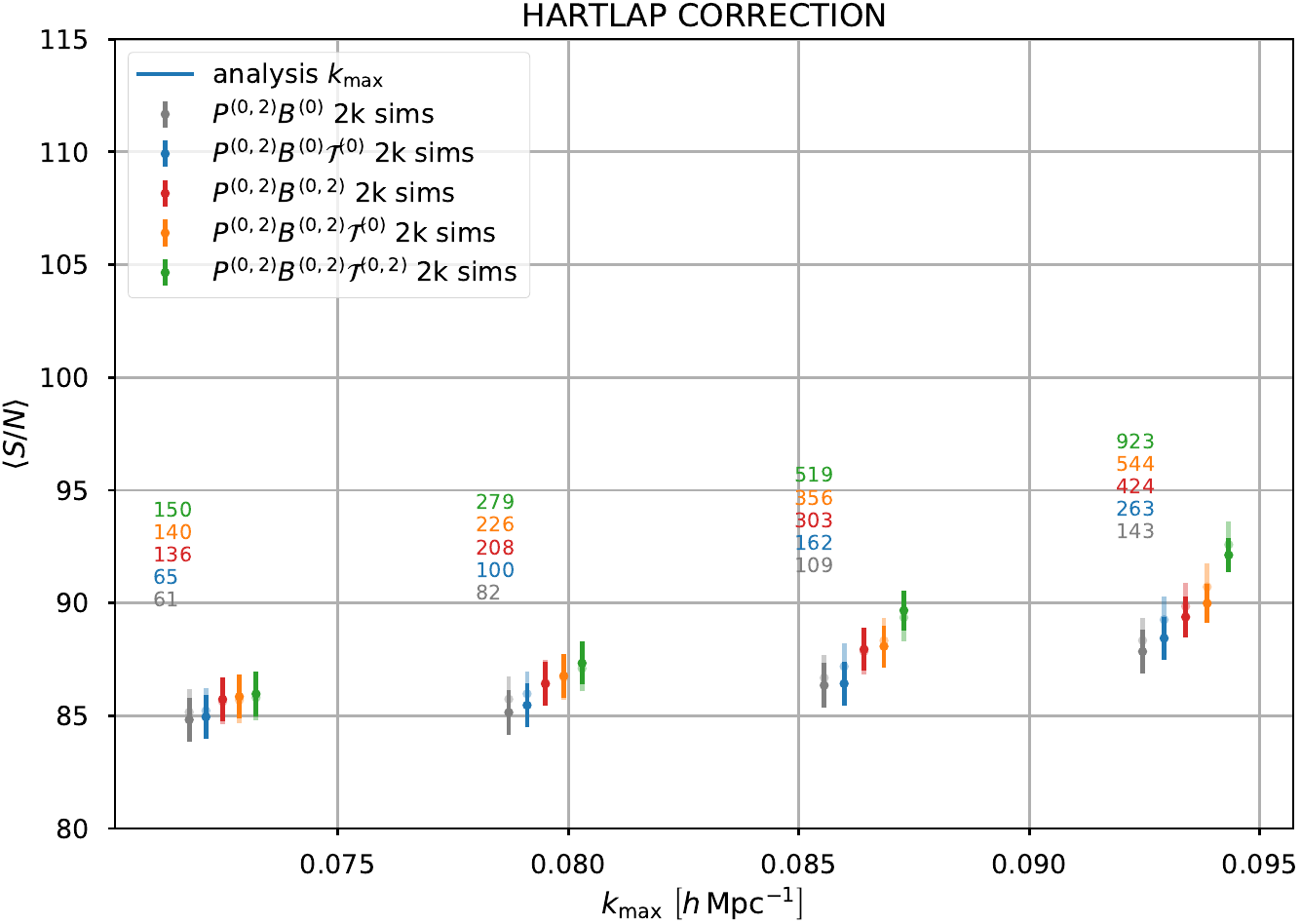}
\caption{
\label{fig:sn_Hartlap} 
Comparison between the $\langle\mr{S/N}\rangle$ computed using 2000 (opaque colours) and 8000 (transparent colours) simulations to estimate the covariance matrices for the different data-vectors combinations, respectively. Applying in both cases the appropriate Hartlap correction the $\langle\mr{S/N}\rangle$ remains approximately constant with respect to the number of simulations used.
}
\end{figure}

We worry that for the highest $k_{\rm max}$ the number of simulations we have does not put us safely in the $n\ll p$ regime. So we estimate the data-vector's covariance using both 2000 and 8000 simulations and apply then the two different corrections to the inverse matrix.
In Fig~ \ref{fig:sn_Hartlap} the dark coloured points are relative to the (S/N) computed from the covariances estimated using only 2000 different realisations. The transparent points correspond to the same quantities but with the covariance estimated from 8000 simulations. Looking at the data-vector containing the trispectrum quadrupoles (green) which corresponds to the largest data-vector case, it is clear that even in this case the number of simulations is sufficient to yield a reliable estimate of the S/N.

\section{Additional material}
\label{app:add_mat}

In Table \ref{tab:table_constr_nuisance} are reported the 1D $68\%$ credible regions for the nuisance parameters together with the relative improvement on constraints due to the addition of bispectrum and i-trispectrum multipoles to the data-vector.

Table \ref{tab:table_constr_fnl0} compares the constraints and relative improvements for both cases of $f_\mr{nl}$ kept constant to zero and for $f_\mr{nl}$ free to vary. 
In terms of constraints improvements, reducing the parameter space, does not affect the impact of adding bispectrum and i-trispectrum multipoles to the data-vector.

\begin{figure}[tbp]
\centering 
\includegraphics[width=1.0\textwidth]
{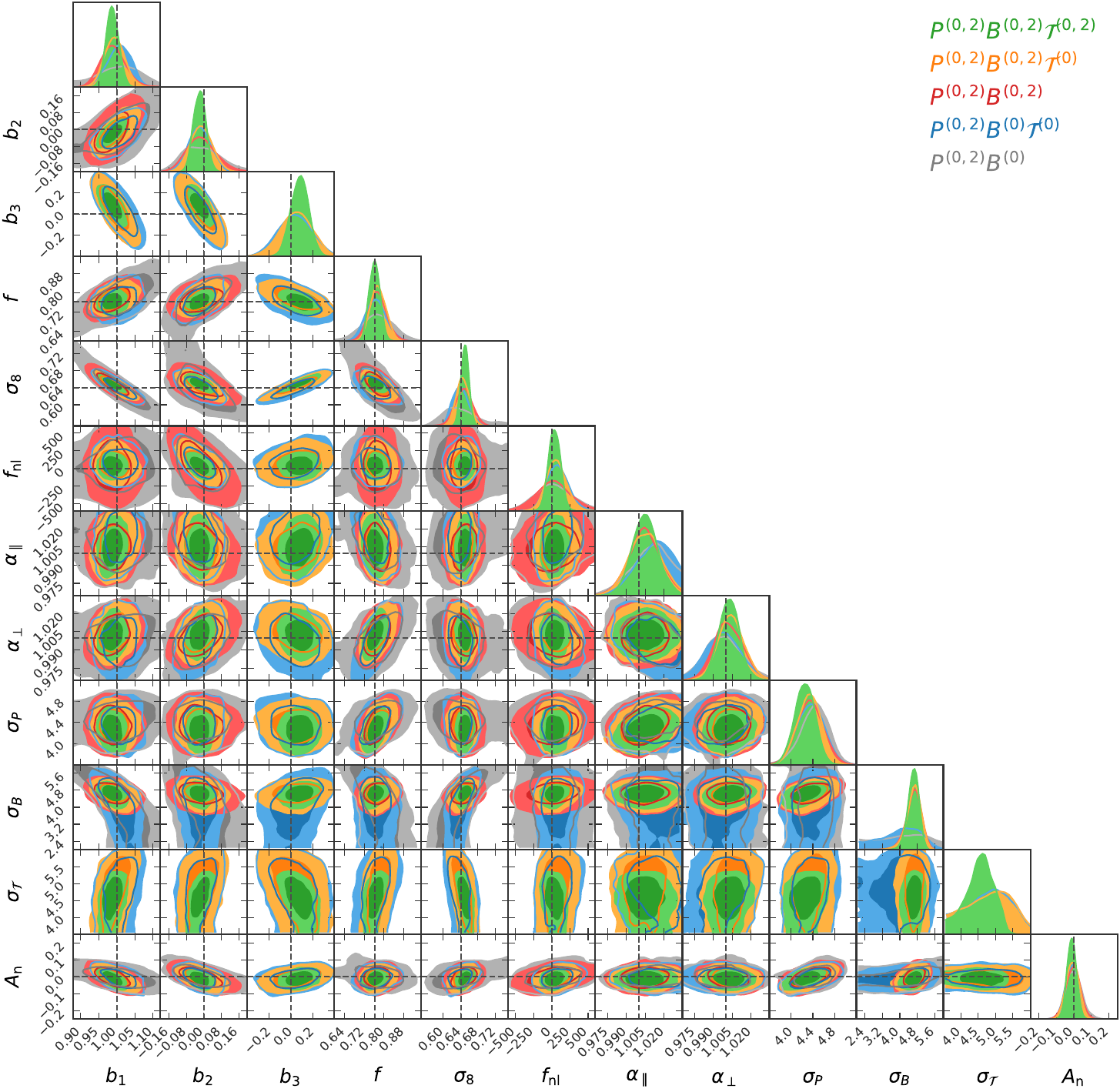}
\caption{\label{fig:mcmc_posteriors}
Same as Figure \ref{fig:cosmo_p_post} but also showing the 1-2D marginalised posterior distributions for the nuisance parameters together with the cosmological ones.
}
\end{figure}

In Figure \ref{fig:mcmc_posteriors} the full parameters set marginalised 1-2D posterior distributions are shown for the case including $f_\mr{nl}$ free to vary.
\renewcommand{\arraystretch}{2.}
\begin{table}[tbp]
\centering
\begin{tabular}{c|cc|cc|cc|cc|cc|}
\cline{2-11}
 & \multicolumn{10}{c|}{MCMC (Fisher Forecasts)} \\
\cline{2-11}
&
 \multicolumn{2}{c|}{$\Delta \theta$}
&
 \multicolumn{8}{c|}{1 - $\left(\Delta\theta/\Delta \theta_{P^{(0,2)}B^{(0)}}\right)$     $\left[\%\right]$}\\
\cline{2-11}
&
\multicolumn{2}{c|}{$P^{(0,2)}B^{(0)}$}  &
\multicolumn{2}{c|}{$P^{(0,2)}B^{(0)}\mathcal{T}^{(0)}$}  & 
\multicolumn{2}{c|}{$P^{(0,2)}B^{(0,2)}$} &
\multicolumn{2}{c|}{$P^{(0,2)}B^{(0,2)}\mathcal{T}^{(0)}$} &
\multicolumn{2}{c|}{$P^{(0,2)}B^{(0,2)}\mathcal{T}^{(0,2)}$} \\
\hline
$b_1$        & \tf{0.128} & (\tf{0.097}) & 48     & (60)     & 38  & (68)  & 57.6 & (73.8) & 74.2 & (80.6) \\
$b_2$        & \tf{0.184} & (\tf{0.122}) & 44.2    & (52)     & 14.7 & (43.8) & 48.7 & (60.2) & 72  & (74.8) \\
$b_3$        &  -    &  -     & \tf{0.3}  & (\tf{0.181}) & -  &  -  & 11.4 & (28.2) & 52.2 & (52.9) \\
$\sigma_P$      & \tf{0.480} & (\tf{0.260}) & 5.3    & (2.5)    & 6.4 & (20.9) & 8  & (22.2) & 25.6 & (29.8) \\
$\sigma_B$      & \tf{2.583} & (\tf{1.533}) & 13.8    & (49.3)    & 72  & (82.5) & 75.6 & (83)  & 79.7 & (85.3) \\
$\sigma_\mathcal{T}$ &  -    &  -     & \tf{1.584} & (\tf{4.414}) & -  &  -  & 9.6 & (9.)  & 55.7 & (63.9) \\
$A_\mr{n}$      & \tf{0.088} & (\tf{0.053}) & 16.3    & (25.4)    & 4.9 & (27.4) & 21.8 & (36.1) & 38.9 & (46.5) \\
\hline
\end{tabular}
\caption{\label{tab:table_constr_nuisance}
Same as Table \ref{tab:table_constr} but reporting the results for the nuisance parameters.
}
\end{table}

\renewcommand{\arraystretch}{2.}
\begin{table}[tbp]
\centering
\begin{tabular}{c|cc|cc|cc|cc|cc|}
\cline{2-11}
 & \multicolumn{10}{c|}{MCMC (Table 1 $f_\mr{nl}$ varying)} \\
\cline{2-11}
&
 \multicolumn{2}{c|}{$\Delta \theta$}
&
 \multicolumn{8}{c|}{1 - $\left(\Delta\theta/\Delta \theta_{P^{(0,2)}B^{(0)}}\right)$     $\left[\%\right]$}\\
\cline{2-11}
&
\multicolumn{2}{c|}{$P^{(0,2)}B^{(0)}$}  &
\multicolumn{2}{c|}{$P^{(0,2)}B^{(0)}\mathcal{T}^{(0)}$}  & 
\multicolumn{2}{c|}{$P^{(0,2)}B^{(0,2)}$} &
\multicolumn{2}{c|}{$P^{(0,2)}B^{(0,2)}\mathcal{T}^{(0)}$} &
\multicolumn{2}{c|}{$P^{(0,2)}B^{(0,2)}\mathcal{T}^{(0,2)}$} \\
\hline
$f$         & \tf{0.124} & (0.140) & 39.1 & (42.8) & 43.3 & (46.4) & 51.1 & (57.7) & 65.4 & (71.9) \\
$\sigma_8$      & \tf{0.072} & (0.078) & 55.5 & (52.3) & 44.2 & (40.6) & 60.1 & (62.3) & 73.3 & (78.2) \\
$\alpha_\parallel$  & \tf{0.035} & (0.036) & 7.8  & (13.9) & 26.9 & (30.6) & 34. & (34.2) & 41.1 & (46.8) \\
$\alpha_\perp$    & \tf{0.030} & (0.032) & 14.1 & (14.3) & 29.5 & (29.1) & 30.8 & (33)  & 44.1 & (46.3) \\
\hline
\multicolumn{3}{c|}{average improvement}
& 29.1 & (30.8) & 36. & (36.7) & 44. & (46.8) & 56 & (60.8) \\
\hline
\end{tabular}
\caption{\label{tab:table_constr_fnl0}
Comparison between 1D $68\%$ credible regions and constraints improvements for the cosmological parameters in both cases of $f_\mr{nl}$ set equal zero or left free to vary (values between brackets). As expected, when reducing the degeneracy in the multi-dimensional posterior distribution by keeping constant $f_\mr{nl}$, the constraints improvement are slightly smaller than the ones for the 12 parameters case. However this reduction is still negligible with respect to the overall improvements.}
\end{table}

\bibliographystyle{ieeetr}
\bibliography{references}

\begin{thebibliography}{100}

\bibitem{Levi:2013gra}
M.~{Levi}, C.~{Bebek}, T.~{Beers}, R.~{Blum}, R.~{Cahn}, D.~{Eisenstein},
  B.~{Flaugher}, K.~{Honscheid}, R.~{Kron}, O.~{Lahav}, P.~{McDonald},
  N.~{Roe}, D.~{Schlegel}, and {representing the DESI collaboration}, ``{The
  DESI Experiment, a whitepaper for Snowmass 2013},'' {\em ArXiv e-prints},
  Aug. 2013.

\bibitem{Laureijs:2011gra}
R.~Laureijs {\em et~al.}, ``{Euclid Definition Study Report},'' 2011.

\bibitem{Ellis:2012rn}
R.~Ellis {\em et~al.}, ``{Extragalactic science, cosmology, and Galactic
  archaeology with the Subaru Prime Focus Spectrograph},'' {\em Publ. Astron.
  Soc. Jap.}, vol.~66, no.~1, p.~R1, 2014.

\bibitem{Bacon:2018dui}
D.~J. Bacon {\em et~al.}, ``{Cosmology with Phase 1 of the Square Kilometre
  Array: Red Book 2018: Technical specifications and performance forecasts},''
  {\em Submitted to: Publ. Astron. Soc. Austral.}, 2018.

\bibitem{Abell:2009aa}
P.~A. Abell {\em et~al.}, ``{LSST Science Book, Version 2.0},'' 2009.

\bibitem{Green:2012mj}
J.~Green {\em et~al.}, ``{Wide-Field InfraRed Survey Telescope (WFIRST) Final
  Report},'' 8 2012.

\bibitem{Peebles1980}
P.~J.~E. {Peebles}, {\em {The large-scale structure of the universe}}.
\newblock 1980.

\bibitem{Groth:1977gj}
E.~J. Groth and P.~J.~E. Peebles, ``{Statistical analysis of catalogs of
  extragalactic objects. 7. Two and three point correlation functions for the
  high - resolution Shane-Wirtanen catalog of galaxies},'' {\em Astrophys. J.},
  vol.~217, p.~385, 1977.

\bibitem{1975ApJ...196....1P}
P.~J.~E. {Peebles} and E.~J. {Groth}, ``{Statistical analysis of catalogs of
  extragalactic objects. V. Three-point correlation function for the galaxy
  distribution in the Zwicky catalog.},'' {\em \apj}, vol.~196, pp.~1--11, Feb.
  1975.

\bibitem{1982ApJ...259..474F}
J.~N. {Fry} and M.~{Seldner}, ``{Transform analysis of the high-resolution
  Shane-Wirtanen Catalog - The power spectrum and the bispectrum},'' {\em
  \apj}, vol.~259, pp.~474--481, Aug. 1982.

\bibitem{Fry:1983cj}
J.~N. Fry, ``{The Galaxy correlation hierarchy in perturbation theory},'' {\em
  Astrophys. J.}, vol.~279, pp.~499--510, 1984.

\bibitem{Matarrese:1997sk}
S.~Matarrese, L.~Verde, and A.~F. Heavens, ``{Large scale bias in the universe:
  Bispectrum method},'' {\em Mon. Not. Roy. Astron. Soc.}, vol.~290,
  pp.~651--662, 1997.

\bibitem{Verde:1998zr}
L.~Verde, A.~F. Heavens, S.~Matarrese, and L.~Moscardini, ``{Large scale bias
  in the universe. 2. Redshift space bispectrum},'' {\em Mon. Not. Roy. Astron.
  Soc.}, vol.~300, pp.~747--756, 1998.

\bibitem{Verde:2001sf}
L.~Verde {\em et~al.}, ``{The 2dF Galaxy Redshift Survey: The Bias of galaxies
  and the density of the Universe},'' {\em Mon. Not. Roy. Astron. Soc.},
  vol.~335, p.~432, 2002.

\bibitem{Scoccimarro:1997st}
R.~Scoccimarro, S.~Colombi, J.~N. Fry, J.~A. Frieman, E.~Hivon, and A.~Melott,
  ``{Nonlinear evolution of the bispectrum of cosmological perturbations},''
  {\em Astrophys. J.}, vol.~496, p.~586, 1998.

\bibitem{Scoccimarro:2000ee}
R.~Scoccimarro and H.~M.~P. Couchman, ``{A fitting formula for the nonlinear
  evolution of the bispectrum},'' {\em Mon. Not. Roy. Astron. Soc.}, vol.~325,
  p.~1312, 2001.

\bibitem{Scoccimarro:1999ed}
R.~Scoccimarro, H.~M.~P. Couchman, and J.~A. Frieman, ``{The Bispectrum as a
  Signature of Gravitational Instability in Redshift-Space},'' {\em Astrophys.
  J.}, vol.~517, pp.~531--540, 1999.

\bibitem{Scoccimarro:2000sp}
R.~Scoccimarro, H.~A. Feldman, J.~N. Fry, and J.~A. Frieman, ``{The Bispectrum
  of IRAS redshift catalogs},'' {\em Astrophys. J.}, vol.~546, p.~652, 2001.

\bibitem{Sefusatti:2009qh}
E.~Sefusatti, ``{1-loop Perturbative Corrections to the Matter and Galaxy
  Bispectrum with non-Gaussian Initial Conditions},'' {\em Phys. Rev. D},
  vol.~80, p.~123002, 2009.

\bibitem{Sefusatti:2011gt}
E.~Sefusatti, M.~Crocce, and V.~Desjacques, ``{The Halo Bispectrum in N-body
  Simulations with non-Gaussian Initial Conditions},'' {\em Mon. Not. Roy.
  Astron. Soc.}, vol.~425, p.~2903, 2012.

\bibitem{Hashimoto:2017klo}
I.~Hashimoto, Y.~Rasera, and A.~Taruya, ``{Precision cosmology with
  redshift-space bispectrum: a perturbation theory based model at one-loop
  order},'' {\em Phys. Rev.}, vol.~D96, no.~4, p.~043526, 2017.

\bibitem{Desjacques:2018pfv}
V.~Desjacques, D.~Jeong, and F.~Schmidt, ``{The Galaxy Power Spectrum and
  Bispectrum in Redshift Space},'' {\em JCAP}, vol.~1812, no.~12, p.~035, 2018.

\bibitem{Eggemeier:2018qae}
A.~Eggemeier, R.~Scoccimarro, and R.~E. Smith, ``{Bias Loop Corrections to the
  Galaxy Bispectrum},'' {\em Phys. Rev.}, vol.~D99, no.~12, p.~123514, 2019.

\bibitem{Castiblanco:2018qsd}
L.~Castiblanco, R.~Gannouji, J.~Noreña, and C.~Stahl, ``{Relativistic
  cosmological large scale structures at one-loop},'' {\em JCAP}, vol.~1907,
  no.~07, p.~030, 2019.

\bibitem{Eggemeier:2021cam}
A.~Eggemeier, R.~Scoccimarro, R.~E. Smith, M.~Crocce, A.~Pezzotta, and A.~G.
  S\'anchez, ``{Testing one-loop galaxy bias: joint analysis of power spectrum
  and bispectrum},'' 2 2021.

\bibitem{Fry:1992vr}
J.~N. Fry and E.~Gaztanaga, ``{Biasing and hierarchical statistics in large
  scale structure},'' {\em Astrophys. J.}, vol.~413, pp.~447--452, 1993.

\bibitem{Fry:1993bj}
J.~N. Fry, ``{The Minimal power spectrum: Higher order contributions},'' {\em
  Astrophys. J.}, vol.~421, pp.~21--26, 1994.

\bibitem{Yankelevich:2018uaz}
V.~Yankelevich and C.~Porciani, ``{Cosmological information in the
  redshift-space bispectrum},'' {\em Mon. Not. Roy. Astron. Soc.}, vol.~483,
  no.~2, pp.~2078--2099, 2019.

\bibitem{Oddo:2019run}
A.~Oddo, E.~Sefusatti, C.~Porciani, P.~Monaco, and A.~G. S\'anchez, ``{Toward a
  robust inference method for the galaxy bispectrum: likelihood function and
  model selection},'' {\em JCAP}, vol.~03, p.~056, 2020.

\bibitem{Barreira:2019icq}
A.~Barreira, ``{The squeezed matter bispectrum covariance with responses},''
  {\em JCAP}, vol.~1903, no.~03, p.~008, 2019.

\bibitem{Agarwal:2020lov}
N.~Agarwal, V.~Desjacques, D.~Jeong, and F.~Schmidt, ``{Information content in
  the redshift-space galaxy power spectrum and bispectrum},'' {\em JCAP},
  vol.~03, p.~021, 2021.

\bibitem{Ruggeri:2017dda}
R.~Ruggeri, E.~Castorina, C.~Carbone, and E.~Sefusatti, ``{DEMNUni: Massive
  neutrinos and the bispectrum of large scale structures},'' {\em JCAP},
  vol.~1803, no.~03, p.~003, 2018.

\bibitem{Coulton:2018ebd}
W.~R. Coulton, J.~Liu, M.~S. Madhavacheril, V.~B\"ohm, and D.~N. Spergel,
  ``{Constraining Neutrino Mass with the Tomographic Weak Lensing
  Bispectrum},'' {\em JCAP}, vol.~05, p.~043, 2019.

\bibitem{Hahn:2019zob}
C.~Hahn, F.~Villaescusa-Navarro, E.~Castorina, and R.~Scoccimarro,
  ``{Constraining $M_\nu$ with the bispectrum. Part I. Breaking parameter
  degeneracies},'' {\em JCAP}, vol.~03, p.~040, 2020.

\bibitem{Hahn:2020lou}
C.~Hahn and F.~Villaescusa-Navarro, ``{Constraining $M_\nu$ with the Bispectrum
  II: The Total Information Content of the Galaxy Bispectrum},'' 12 2020.

\bibitem{Kamalinejad:2020izi}
F.~Kamalinejad and Z.~Slepian, ``{A Non-Degenerate Neutrino Mass Signature in
  the Galaxy Bispectrum},'' 11 2020.

\bibitem{GilMarin:2011xq}
H.~Gil-Marin, F.~Schmidt, W.~Hu, R.~Jimenez, and L.~Verde, ``{The Bispectrum of
  f(R) Cosmologies},'' {\em JCAP}, vol.~1111, p.~019, 2011.

\bibitem{Bartolo:2013ws}
N.~Bartolo, E.~Bellini, D.~Bertacca, and S.~Matarrese, ``{Matter bispectrum in
  cubic Galileon cosmologies},'' {\em JCAP}, vol.~1303, p.~034, 2013.

\bibitem{Bellini:2015wfa}
E.~Bellini, R.~Jimenez, and L.~Verde, ``{Signatures of Horndeski gravity on the
  Dark Matter Bispectrum},'' {\em JCAP}, vol.~1505, no.~05, p.~057, 2015.

\bibitem{Bertacca:2017dzm}
D.~Bertacca, A.~Raccanelli, N.~Bartolo, M.~Liguori, S.~Matarrese, and L.~Verde,
  ``{Relativistic wide-angle galaxy bispectrum on the light-cone},'' {\em Phys.
  Rev.}, vol.~D97, no.~2, p.~023531, 2018.

\bibitem{DiDio:2018unb}
E.~Di~Dio, R.~Durrer, R.~Maartens, F.~Montanari, and O.~Umeh, ``{The Full-Sky
  Angular Bispectrum in Redshift Space},'' {\em JCAP}, vol.~04, p.~053, 2019.

\bibitem{Pearson:2017wtw}
D.~W. Pearson and L.~Samushia, ``{A Detection of the Baryon Acoustic
  Oscillation features in the SDSS BOSS DR12 Galaxy Bispectrum},'' {\em Mon.
  Not. Roy. Astron. Soc.}, vol.~478, no.~4, pp.~4500--4512, 2018.

\bibitem{Child:2018klv}
H.~L. Child, M.~Takada, T.~Nishimichi, T.~Sunayama, Z.~Slepian, S.~Habib, and
  K.~Heitmann, ``{Bispectrum as Baryon Acoustic Oscillation Interferometer},''
  {\em Phys. Rev. D}, vol.~98, no.~12, p.~123521, 2018.

\bibitem{Slepian:2015hca}
Z.~Slepian {\em et~al.}, ``{The large-scale three-point correlation function of
  the SDSS BOSS DR12 CMASS galaxies},'' {\em Mon. Not. Roy. Astron. Soc.},
  vol.~468, no.~1, pp.~1070--1083, 2017.

\bibitem{Slepian:2016kfz}
Z.~Slepian {\em et~al.}, ``{Detection of baryon acoustic oscillation features
  in the large-scale three-point correlation function of SDSS BOSS DR12 CMASS
  galaxies},'' {\em Mon. Not. Roy. Astron. Soc.}, vol.~469, no.~2,
  pp.~1738--1751, 2017.

\bibitem{Verde:1999ij}
L.~Verde, L.-M. Wang, A.~Heavens, and M.~Kamionkowski, ``{Large scale
  structure, the cosmic microwave background, and primordial
  non-gaussianity},'' {\em Mon. Not. Roy. Astron. Soc.}, vol.~313,
  pp.~L141--L147, 2000.

\bibitem{Scoccimarro:2003wn}
R.~Scoccimarro, E.~Sefusatti, and M.~Zaldarriaga, ``{Probing primordial
  non-Gaussianity with large - scale structure},'' {\em Phys. Rev.}, vol.~D69,
  p.~103513, 2004.

\bibitem{Jeong:2009vd}
D.~Jeong and E.~Komatsu, ``{Primordial non-Gaussianity, scale-dependent bias,
  and the bispectrum of galaxies},'' {\em Astrophys. J.}, vol.~703,
  pp.~1230--1248, 2009.

\bibitem{Bose:2018zpk}
B.~Bose and A.~Taruya, ``{The one-loop matter bispectrum as a probe of gravity
  and dark energy},'' {\em JCAP}, vol.~1810, no.~10, p.~019, 2018.

\bibitem{Karagiannis:2018jdt}
D.~Karagiannis, A.~Lazanu, M.~Liguori, A.~Raccanelli, N.~Bartolo, and L.~Verde,
  ``{Constraining primordial non-Gaussianity with bispectrum and power spectrum
  from upcoming optical and radio surveys},'' {\em Mon. Not. Roy. Astron.
  Soc.}, vol.~478, no.~1, pp.~1341--1376, 2018.

\bibitem{Takada:2003ef}
M.~Takada and B.~Jain, ``{Cosmological parameters from lensing power spectrum
  and bispectrum tomography},'' {\em Mon. Not. Roy. Astron. Soc.}, vol.~348,
  p.~897, 2004.

\bibitem{Kilbinger:2005jy}
M.~Kilbinger and P.~Schneider, ``{Cosmological parameters from combined second-
  and third-order aperture mass statistics of cosmic shear},'' {\em Astron.
  Astrophys.}, vol.~442, p.~69, 2005.

\bibitem{Semboloni:2010er}
E.~Semboloni, T.~Schrabback, L.~van Waerbeke, S.~Vafaei, J.~Hartlap, and
  S.~Hilbert, ``{Weak lensing from space: first cosmological constraints from
  three-point shear statistics},'' {\em Mon. Not. Roy. Astron. Soc.}, vol.~410,
  p.~143, 2011.

\bibitem{Kayo:2012nm}
I.~Kayo, M.~Takada, and B.~Jain, ``{Information content of weak lensing power
  spectrum and bispectrum: including the non-Gaussian error covariance
  matrix},'' {\em Mon. Not. Roy. Astron. Soc.}, vol.~429, pp.~344--371, 2013.

\bibitem{Rizzato:2018whp}
M.~Rizzato, K.~Benabed, F.~Bernardeau, and F.~Lacasa, ``{Tomographic weak
  lensing bispectrum: a thorough analysis towards the next generation of galaxy
  surveys},'' {\em Mon. Not. Roy. Astron. Soc.}, vol.~490, pp.~4688--4714,
  2019.

\bibitem{Halder:2021itp}
A.~Halder, O.~Friedrich, S.~Seitz, and T.~N. Varga, ``{The integrated 3-point
  correlation function of cosmic shear},'' 2 2021.

\bibitem{Jung:2021weh}
G.~Jung, T.~Namikawa, M.~Liguori, D.~Munshi, and A.~Heavens, ``{The integrated
  angular bispectrum of weak lensing},'' 2 2021.

\bibitem{Pyne:2020ijd}
S.~Pyne and B.~Joachimi, ``{Self-calibration of weak lensing systematic effects
  using combined two- and three-point statistics},'' {\em Mon. Not. Roy.
  Astron. Soc.}, vol.~503, no.~2, pp.~2300--2317, 2021.

\bibitem{Fu:2014loa}
L.~Fu {\em et~al.}, ``{CFHTLenS: Cosmological constraints from a combination of
  cosmic shear two-point and three-point correlations},'' {\em Mon. Not. Roy.
  Astron. Soc.}, vol.~441, pp.~2725--2743, 2014.

\bibitem{Gil-Marin:2014sta}
H.~Gil-Marín, J.~Noreña, L.~Verde, W.~J. Percival, C.~Wagner, M.~Manera, and
  D.~P. Schneider, ``{The power spectrum and bispectrum of SDSS DR11 BOSS
  galaxies – I. Bias and gravity},'' {\em Mon. Not. Roy. Astron. Soc.},
  vol.~451, no.~1, pp.~539--580, 2015.

\bibitem{Gil-Marin:2016wya}
H.~Gil-Marín, W.~J. Percival, L.~Verde, J.~R. Brownstein, C.-H. Chuang, F.-S.
  Kitaura, S.~A. Rodríguez-Torres, and M.~D. Olmstead, ``{The clustering of
  galaxies in the SDSS-III Baryon Oscillation Spectroscopic Survey: RSD
  measurement from the power spectrum and bispectrum of the DR12 BOSS
  galaxies},'' {\em Mon. Not. Roy. Astron. Soc.}, vol.~465, no.~2,
  pp.~1757--1788, 2017.

\bibitem{Scoccimarro:2015bla}
R.~Scoccimarro, ``{Fast Estimators for Redshift-Space Clustering},'' {\em Phys.
  Rev. D}, vol.~92, no.~8, p.~083532, 2015.

\bibitem{Sugiyama:2018yzo}
N.~S. Sugiyama, S.~Saito, F.~Beutler, and H.-J. Seo, ``{A complete FFT-based
  decomposition formalism for the redshift-space bispectrum},'' {\em Mon. Not.
  Roy. Astron. Soc.}, vol.~484, no.~1, pp.~364--384, 2019.

\bibitem{Sugiyama:2019ike}
N.~S. Sugiyama, S.~Saito, F.~Beutler, and H.-J. Seo, ``{Perturbation theory
  approach to predict the covariance matrices of the galaxy power spectrum and
  bispectrum in redshift space},'' {\em Mon. Not. Roy. Astron. Soc.}, vol.~497,
  no.~2, pp.~1684--1711, 2020.

\bibitem{Gagrani:2016rfy}
P.~Gagrani and L.~Samushia, ``{Information Content of the Angular Multipoles of
  Redshift-Space Galaxy Bispectrum},'' {\em Mon. Not. Roy. Astron. Soc.},
  vol.~467, no.~1, pp.~928--935, 2017.

\bibitem{Gualdi:2020ymf}
D.~Gualdi and L.~Verde, ``{Galaxy redshift-space bispectrum: the Importance of
  Being Anisotropic},'' {\em JCAP}, vol.~06, p.~041, 2020.

\bibitem{kunzetal2001}
M.~{Kunz}, A.~J. {Banday}, P.~G. {Castro}, P.~G. {Ferreira}, and K.~M.
  {G{\'o}rski}, ``{The Trispectrum of the 4 Year COBE DMR Data},'' {\em \apjl},
  vol.~563, pp.~L99--L102, Dec. 2001.

\bibitem{Komatsu:2002db}
E.~Komatsu, {\em {The pursuit of non-gaussian fluctuations in the cosmic
  microwave background}}.
\newblock PhD thesis, Tohoku U., 2001.

\bibitem{deTroia2003}
G.~de~Troia, P.~A.~R. Ade, J.~J. Bock, J.~R. Bond, A.~Boscaleri, C.~R.
  Contaldi, B.~P. Crill, P.~de~Bernardis, P.~G. Ferreira, M.~Giacometti,
  E.~Hivon, V.~V. Hristov, M.~Kunz, A.~E. Lange, S.~Masi, P.~D. Mauskopf,
  T.~Montroy, P.~Natoli, C.~B. Netterfield, E.~Pascale, F.~Piacentini,
  G.~Polenta, G.~Romeo, and J.~E. Ruhl, ``{The trispectrum of the cosmic
  microwave background on subdegree angular scales: an analysis of the
  BOOMERanG data},'' {\em Monthly Notices of the Royal Astronomical Society},
  vol.~343, pp.~284--292, 07 2003.

\bibitem{Munshi:2009wy}
D.~Munshi, A.~Heavens, A.~Cooray, J.~Smidt, P.~Coles, and P.~Serra, ``{New
  Optimised Estimators for the Primordial Trispectrum},'' {\em Mon. Not. Roy.
  Astron. Soc.}, vol.~412, p.~1993, 2011.

\bibitem{Kamionkowski:2010me}
M.~Kamionkowski, T.~L. Smith, and A.~Heavens, ``{The CMB Bispectrum,
  Trispectrum, non-Gaussianity, and the Cramer-Rao Bound},'' {\em Phys. Rev.
  D}, vol.~83, p.~023007, 2011.

\bibitem{Izumi:2011di}
K.~Izumi, S.~Mizuno, and K.~Koyama, ``{Trispectrum estimation in various models
  of equilateral type non-Gaussianity},'' {\em Phys. Rev. D}, vol.~85,
  p.~023521, 2012.

\bibitem{Regan_2015}
D.~Regan, M.~Gosenca, and D.~Seery, ``Constraining the {WMAP}9 bispectrum and
  trispectrum with needlets,'' {\em Journal of Cosmology and Astroparticle
  Physics}, vol.~2015, pp.~013--013, jan 2015.

\bibitem{Feng:2015pva}
C.~Feng, A.~Cooray, J.~Smidt, J.~O'Bryan, B.~Keating, and D.~Regan, ``{Planck
  Trispectrum Constraints on Primordial Non-Gaussianity at Cubic Order},'' {\em
  Phys. Rev. D}, vol.~92, no.~4, p.~043509, 2015.

\bibitem{Fergusson:2010gn}
J.~Fergusson, D.~Regan, and E.~Shellard, ``{Optimal Trispectrum Estimators and
  WMAP Constraints},'' 12 2010.

\bibitem{Smith:2015uia}
K.~M. Smith, L.~Senatore, and M.~Zaldarriaga, ``{Optimal analysis of the CMB
  trispectrum},'' 2 2015.

\bibitem{Namikawa:2017uke}
T.~Namikawa, ``{Constraints on Patchy Reionization from Planck CMB Temperature
  Trispectrum},'' {\em Phys. Rev. D}, vol.~97, no.~6, p.~063505, 2018.

\bibitem{PlanckTrispectrum18}
T.~{Namikawa}, ``{Constraints on patchy reionization from Planck CMB
  temperature trispectrum},'' {\em \prd}, vol.~97, p.~063505, Mar. 2018.

\bibitem{Akrami:2019izv}
Y.~Akrami {\em et~al.}, ``{Planck 2018 results. IX. Constraints on primordial
  non-Gaussianity},'' {\em Astron. Astrophys.}, vol.~641, p.~A9, 2020.

\bibitem{Verde:2001pf}
L.~Verde and A.~F. Heavens, ``{On the trispectrum as a Gaussian test for
  cosmology},'' {\em Astrophys. J.}, vol.~553, p.~14, 2001.

\bibitem{Cooray:2008eb}
A.~Cooray, C.~Li, and A.~Melchiorri, ``{The trispectrum of 21-cm background
  anisotropies as a probe of primordial non-Gaussianity},'' {\em Phys. Rev. D},
  vol.~77, p.~103506, 2008.

\bibitem{Lazeyras:2017hxw}
T.~Lazeyras and F.~Schmidt, ``{Beyond LIMD bias: a measurement of the complete
  set of third-order halo bias parameters},'' {\em JCAP}, vol.~09, p.~008,
  2018.

\bibitem{Bellomo:2018lew}
N.~Bellomo, N.~Bartolo, R.~Jimenez, S.~Matarrese, and L.~Verde, ``{Measuring
  the Energy Scale of Inflation with Large Scale Structures},'' {\em JCAP},
  vol.~11, p.~043, 2018.

\bibitem{Fry1978}
J.~N. {Fry} and P.~J.~E. {Peebles}, ``{Statistical analysis of catalogs of
  extragalactic objects. IX. The four-point galaxy correlation function.},''
  {\em \apj}, vol.~221, pp.~19--33, Apr. 1978.

\bibitem{Suto:1993ua}
Y.~Suto and T.~Matsubara, ``{Departure from hierarchical clustering relations
  for two, three, and four point correlation functions: Analysis of
  cosmological N body simulations},'' 3 1993.

\bibitem{Sabiu:2019kbh}
C.~G. Sabiu, B.~Hoyle, J.~Kim, and X.-D. Li, ``{Graph Database Solution for
  Higher Order Spatial Statistics in the Era of Big Data},'' {\em Astrophys. J.
  Suppl.}, vol.~242, no.~2, p.~29, 2019.

\bibitem{2010A&A...514A..79P}
J.~{Pielorz}, J.~{R{\"o}diger}, I.~{Tereno}, and P.~{Schneider}, ``{A fitting
  formula for the non-Gaussian contribution to the lensing power spectrum
  covariance},'' {\em \aap}, vol.~514, p.~A79, May 2010.

\bibitem{2012JCAP...04..019D}
R.~{de Putter}, C.~{Wagner}, O.~{Mena}, L.~{Verde}, and W.~J. {Percival},
  ``{Thinking outside the box: effects of modes larger than the survey on
  matter power spectrum covariance},'' {\em \jcap}, vol.~2012, p.~019, Apr.
  2012.

\bibitem{Mohammed:2016sre}
I.~Mohammed, U.~Seljak, and Z.~Vlah, ``{Perturbative approach to covariance
  matrix of the matter power spectrum},'' {\em Mon. Not. Roy. Astron. Soc.},
  vol.~466, no.~1, pp.~780--797, 2017.

\bibitem{Taruya:2020qoy}
A.~Taruya, T.~Nishimichi, and D.~Jeong, ``{Covariance of the matter power
  spectrum including the survey window function effect: $N$ -body simulations
  versus fifth-order perturbation theory on grids},'' {\em Phys. Rev. D},
  vol.~103, no.~2, p.~023501, 2021.

\bibitem{Bertolini:2016bmt}
D.~Bertolini, K.~Schutz, M.~P. Solon, and K.~M. Zurek, ``{The Trispectrum in
  the Effective Field Theory of Large Scale Structure},'' {\em JCAP}, vol.~06,
  p.~052, 2016.

\bibitem{Steele:2021lnz}
T.~Steele and T.~Baldauf, ``{Precise Calibration of the One-Loop Trispectrum in
  the Effective Field Theory of Large Scale Structure},'' 1 2021.

\bibitem{Lee:2020ebj}
H.~Lee and C.~Dvorkin, ``{Cosmological Angular Trispectra and Non-Gaussian
  Covariance},'' {\em JCAP}, vol.~05, p.~044, 2020.

\bibitem{Lazeyras:2015giz}
T.~Lazeyras, M.~Musso, and V.~Desjacques, ``{Lagrangian bias of generic
  large-scale structure tracers},'' {\em Phys. Rev. D}, vol.~93, no.~6,
  p.~063007, 2016.

\bibitem{Bartolo:2004if}
N.~Bartolo, E.~Komatsu, S.~Matarrese, and A.~Riotto, ``{Non-Gaussianity from
  inflation: Theory and observations},'' {\em Phys. Rept.}, vol.~402,
  pp.~103--266, 2004.

\bibitem{Sefusatti:2004xz}
E.~Sefusatti and R.~Scoccimarro, ``{Galaxy bias and halo-occupation numbers
  from large-scale clustering},'' {\em Phys. Rev. D}, vol.~71, p.~063001, 2005.

\bibitem{Gualdi:2020eag}
D.~Gualdi, S.~Novell, H.~Gil-Mar\'\i{}n, and L.~Verde, ``{Matter trispectrum:
  theoretical modelling and comparison to N-body simulations},'' {\em JCAP},
  vol.~01, p.~015, 2021.

\bibitem{Villaescusa-Navarro:2019bje}
F.~Villaescusa-Navarro {\em et~al.}, ``{The Quijote simulations},'' {\em
  Astrophys. J. Suppl.}, vol.~250, no.~1, p.~2, 2020.

\bibitem{Aghamousa:2016zmz}
A.~Aghamousa {\em et~al.}, ``{The DESI Experiment Part I: Science,Targeting,
  and Survey Design},'' 10 2016.

\bibitem{2013PASP..125..306F}
D.~{Foreman-Mackey}, D.~W. {Hogg}, D.~{Lang}, and J.~{Goodman}, ``{emcee: The
  MCMC Hammer},'' {\em Publications of the Astronomical Society of the
  Pacific}, vol.~125, p.~306, Mar. 2013.

\bibitem{Springel:2002uv}
V.~Springel and L.~Hernquist, ``{Cosmological SPH simulations: A Hybrid
  multi-phase model for star formation},'' {\em Mon. Not. Roy. Astron. Soc.},
  vol.~339, p.~289, 2003.

\bibitem{Crocce:2006ve}
M.~Crocce, S.~Pueblas, and R.~Scoccimarro, ``{Transients from Initial
  Conditions in Cosmological Simulations},'' {\em Mon. Not. Roy. Astron. Soc.},
  vol.~373, pp.~369--381, 2006.

\bibitem{Scoccimarro:2011pz}
R.~Scoccimarro, L.~Hui, M.~Manera, and K.~C. Chan, ``{Large-scale Bias and
  Efficient Generation of Initial Conditions for Non-Local Primordial
  Non-Gaussianity},'' {\em Phys. Rev. D}, vol.~85, p.~083002, 2012.

\bibitem{collaboration2018planck}
P.~Collaboration, ``Planck 2018 results. vi. cosmological parameters,'' 2018.

\bibitem{Taruya:2010mx}
A.~Taruya, T.~Nishimichi, and S.~Saito, ``{Baryon Acoustic Oscillations in 2D:
  Modeling Redshift-space Power Spectrum from Perturbation Theory},'' {\em
  Phys. Rev. D}, vol.~82, p.~063522, 2010.

\bibitem{Nishimichi:2011jm}
T.~Nishimichi and A.~Taruya, ``{Baryon Acoustic Oscillations in 2D II:
  Redshift-space halo clustering in N-body simulations},'' {\em Phys. Rev. D},
  vol.~84, p.~043526, 2011.

\bibitem{Bernardeau:2001qr}
F.~Bernardeau, S.~Colombi, E.~Gaztanaga, and R.~Scoccimarro, ``{Large scale
  structure of the universe and cosmological perturbation theory},'' {\em Phys.
  Rept.}, vol.~367, pp.~1--248, 2002.

\bibitem{Crocce:2005xy}
M.~Crocce and R.~Scoccimarro, ``{Renormalized cosmological perturbation
  theory},'' {\em Phys. Rev. D}, vol.~73, p.~063519, 2006.

\bibitem{Lesgourgues:2011re}
J.~Lesgourgues, ``{The Cosmic Linear Anisotropy Solving System (CLASS) I:
  Overview},'' 4 2011.

\bibitem{GilMarin:2011ik}
H.~Gil-Marin, C.~Wagner, F.~Fragkoudi, R.~Jimenez, and L.~Verde, ``{An improved
  fitting formula for the dark matter bispectrum},'' {\em JCAP}, vol.~02,
  p.~047, 2012.

\bibitem{Baumann:2010tm}
D.~Baumann, A.~Nicolis, L.~Senatore, and M.~Zaldarriaga, ``{Cosmological
  Non-Linearities as an Effective Fluid},'' {\em JCAP}, vol.~07, p.~051, 2012.

\bibitem{Carrasco:2012cv}
J.~J.~M. Carrasco, M.~P. Hertzberg, and L.~Senatore, ``{The Effective Field
  Theory of Cosmological Large Scale Structures},'' {\em JHEP}, vol.~09,
  p.~082, 2012.

\bibitem{Carrasco:2013mua}
J.~J.~M. Carrasco, S.~Foreman, D.~Green, and L.~Senatore, ``{The Effective
  Field Theory of Large Scale Structures at Two Loops},'' {\em JCAP}, vol.~07,
  p.~057, 2014.

\bibitem{Pajer:2013jj}
E.~Pajer and M.~Zaldarriaga, ``{On the Renormalization of the Effective Field
  Theory of Large Scale Structures},'' {\em JCAP}, vol.~08, p.~037, 2013.

\bibitem{Mercolli:2013bsa}
L.~Mercolli and E.~Pajer, ``{On the velocity in the Effective Field Theory of
  Large Scale Structures},'' {\em JCAP}, vol.~03, p.~006, 2014.

\bibitem{Baldauf:2014qfa}
T.~Baldauf, L.~Mercolli, M.~Mirbabayi, and E.~Pajer, ``{The Bispectrum in the
  Effective Field Theory of Large Scale Structure},'' {\em JCAP}, vol.~05,
  p.~007, 2015.

\bibitem{Angulo:2014tfa}
R.~E. Angulo, S.~Foreman, M.~Schmittfull, and L.~Senatore, ``{The One-Loop
  Matter Bispectrum in the Effective Field Theory of Large Scale Structures},''
  {\em JCAP}, vol.~10, p.~039, 2015.

\bibitem{Steele:2020tak}
T.~Steele and T.~Baldauf, ``{Precise Calibration of the One-Loop Bispectrum in
  the Effective Field Theory of Large Scale Structure},'' {\em Phys. Rev. D},
  vol.~103, no.~2, p.~023520, 2021.

\bibitem{Komatsu:2001rj}
E.~Komatsu and D.~N. Spergel, ``{Acoustic signatures in the primary microwave
  background bispectrum},'' {\em Phys. Rev.}, vol.~D63, p.~063002, 2001.

\bibitem{Dalal:2007cu}
N.~Dalal, O.~Dore, D.~Huterer, and A.~Shirokov, ``{The imprints of primordial
  non-gaussianities on large-scale structure: scale dependent bias and
  abundance of virialized objects},'' {\em Phys. Rev. D}, vol.~77, p.~123514,
  2008.

\bibitem{Matarrese:2008nc}
S.~Matarrese and L.~Verde, ``{The effect of primordial non-Gaussianity on halo
  bias},'' {\em Astrophys. J. Lett.}, vol.~677, pp.~L77--L80, 2008.

\bibitem{Giannantonio:2009ak}
T.~Giannantonio and C.~Porciani, ``{Structure formation from non-Gaussian
  initial conditions: multivariate biasing, statistics, and comparison with
  N-body simulations},'' {\em Phys. Rev. D}, vol.~81, p.~063530, 2010.

\bibitem{Baldauf:2010vn}
T.~Baldauf, U.~Seljak, and L.~Senatore, ``{Primordial non-Gaussianity in the
  Bispectrum of the Halo Density Field},'' {\em JCAP}, vol.~1104, p.~006, 2011.

\bibitem{Tellarini:2015faa}
M.~Tellarini, A.~J. Ross, G.~Tasinato, and D.~Wands, ``{Non-local bias in the
  halo bispectrum with primordial non-Gaussianity},'' {\em JCAP}, vol.~07,
  p.~004, 2015.

\bibitem{Abidi:2018eyd}
M.~M. Abidi and T.~Baldauf, ``{Cubic Halo Bias in Eulerian and Lagrangian
  Space},'' {\em JCAP}, vol.~07, p.~029, 2018.

\bibitem{Chan:2012jj}
K.~C. Chan, R.~Scoccimarro, and R.~K. Sheth, ``{Gravity and Large-Scale
  Non-local Bias},'' {\em Phys. Rev. D}, vol.~85, p.~083509, 2012.

\bibitem{Baldauf:2012hs}
T.~Baldauf, U.~Seljak, V.~Desjacques, and P.~McDonald, ``{Evidence for
  Quadratic Tidal Tensor Bias from the Halo Bispectrum},'' {\em Phys. Rev. D},
  vol.~86, p.~083540, 2012.

\bibitem{Saito:2014qha}
S.~Saito, T.~Baldauf, Z.~Vlah, U.~Seljak, T.~Okumura, and P.~McDonald,
  ``{Understanding higher-order nonlocal halo bias at large scales by combining
  the power spectrum with the bispectrum},'' {\em Phys. Rev. D}, vol.~90,
  no.~12, p.~123522, 2014.

\bibitem{Alcock:1979mp}
C.~Alcock and B.~Paczynski, ``{An evolution free test for non-zero cosmological
  constant},'' {\em Nature}, vol.~281, pp.~358--359, 1979.

\bibitem{Brieden:2020upf}
S.~Brieden, H.~Gil-Mar\'\i{}n, L.~Verde, and J.~L. Bernal, ``{Blind Observers
  of the Sky},'' {\em JCAP}, vol.~09, p.~052, 2020.

\bibitem{Brieden:2021edu}
S.~Brieden, H.~Gil-Mar\'\i{}n, and L.~Verde, ``{ShapeFit: Extracting the power
  spectrum shape information in galaxy surveys beyond BAO and RSD},'' 6 2021.

\bibitem{Beutler:2016arn}
F.~Beutler {\em et~al.}, ``{The clustering of galaxies in the completed
  SDSS-III Baryon Oscillation Spectroscopic Survey: Anisotropic galaxy
  clustering in Fourier-space},'' {\em Mon. Not. Roy. Astron. Soc.}, vol.~466,
  no.~2, pp.~2242--2260, 2017.

\bibitem{Gil-Marin:2014pva}
H.~Gil-Marín, C.~Wagner, J.~Noreña, L.~Verde, and W.~Percival, ``{Dark matter
  and halo bispectrum in redshift space: theory and applications},'' {\em
  JCAP}, vol.~12, p.~029, 2014.

\bibitem{2010PhDT.........4J}
D.~{Jeong}, {\em {Cosmology with high (z>1) redshift galaxy surveys}}.
\newblock PhD thesis, University of Texas at Austin, Aug. 2010.

\bibitem{Sefusatti:2015aex}
E.~Sefusatti, M.~Crocce, R.~Scoccimarro, and H.~Couchman, ``{Accurate
  Estimators of Correlation Functions in Fourier Space},'' {\em Mon. Not. Roy.
  Astron. Soc.}, vol.~460, no.~4, pp.~3624--3636, 2016.

\bibitem{10.1145/1464291.1464352}
W.~M. Gentleman and G.~Sande, ``Fast fourier transforms: For fun and profit,''
  in {\em Proceedings of the November 7-10, 1966, Fall Joint Computer
  Conference}, AFIPS ’66 (Fall), (New York, NY, USA), p.~563–578,
  Association for Computing Machinery, 1966.

\bibitem{Scoccimarro:2000sn}
R.~Scoccimarro, ``{The bispectrum: from theory to observations},'' {\em
  Astrophys. J.}, vol.~544, p.~597, 2000.

\bibitem{Tomlinson:2019bjx}
J.~Tomlinson, D.~Jeong, and J.~Kim, ``{Efficient parallel algorithm for
  estimating higher-order polyspectra},'' {\em Astron.\ J.}, vol.~158, no.~3,
  p.~116, 2019.

\bibitem{Tegmark:1996bz}
M.~Tegmark, A.~Taylor, and A.~Heavens, ``{Karhunen-Loeve eigenvalue problems in
  cosmology: How should we tackle large data sets?},'' {\em Astrophys. J.},
  vol.~480, p.~22, 1997.

\bibitem{Carron:2012pw}
J.~Carron, ``{On the assumption of Gaussianity for cosmological two-point
  statistics and parameter dependent covariance matrices},'' {\em Astron.
  Astrophys.}, vol.~551, p.~A88, 2013.

\bibitem{Kalus:2015lna}
B.~Kalus, W.~Percival, and L.~Samushia, ``{Cosmological parameter inference
  from galaxy clustering: The effect of the posterior distribution of the power
  spectrum},'' {\em Mon. Not. Roy. Astron. Soc.}, vol.~455, no.~3,
  pp.~2573--2581, 2016.

\bibitem{Hartlap:2006kj}
J.~Hartlap, P.~Simon, and P.~Schneider, ``{Why your model parameter confidences
  might be too optimistic: Unbiased estimation of the inverse covariance
  matrix},'' {\em Astron. Astrophys.}, vol.~464, p.~399, 2007.

\bibitem{Sellentin:2015waz}
E.~Sellentin and A.~F. Heavens, ``{Parameter inference with estimated
  covariance matrices},'' {\em Mon. Not. Roy. Astron. Soc.}, vol.~456, no.~1,
  pp.~L132--L136, 2016.

\bibitem{Reid:2010vc}
B.~A. Reid, L.~Verde, K.~Dolag, S.~Matarrese, and L.~Moscardini,
  ``{Non-Gaussian halo assembly bias},'' {\em JCAP}, vol.~07, p.~013, 2010.

\bibitem{Lazanu:2015rta}
A.~Lazanu, T.~Giannantonio, M.~Schmittfull, and E.~P.~S. Shellard, ``{Matter
  bispectrum of large-scale structure: Three-dimensional comparison between
  theoretical models and numerical simulations},'' {\em Phys. Rev. D}, vol.~93,
  no.~8, p.~083517, 2016.

\bibitem{Novell:2021}
S.~Novell and {et al.}, ``{in prep.},''

\bibitem{Colless:2001gk}
M.~Colless {\em et~al.}, ``{The 2dF Galaxy Redshift Survey: Spectra and
  redshifts},'' {\em Mon. Not. Roy. Astron. Soc.}, vol.~328, p.~1039, 2001.

\bibitem{2010MNRAS.401.1429D}
M.~J. {Drinkwater}, R.~J. {Jurek}, and e.~a. {Blake}, ``{The WiggleZ Dark
  Energy Survey: survey design and first data release},'' {\em \mnras},
  vol.~401, pp.~1429--1452, Jan. 2010.

\bibitem{2013AJ....145...10D}
K.~S. {Dawson}, D.~J. {Schlegel}, and e.~a. {Ahn}, ``{The Baryon Oscillation
  Spectroscopic Survey of SDSS-III},'' {\em \aj}, vol.~145, p.~10, Jan. 2013.

\bibitem{Sugiyama:2020uil}
N.~S. Sugiyama, S.~Saito, F.~Beutler, and H.-J. Seo, ``{Towards a
  self-consistent analysis of the anisotropic galaxy two- and three-point
  correlation functions on large scales: application to mock galaxy
  catalogues},'' {\em Mon. Not. Roy. Astron. Soc.}, vol.~501, no.~2,
  pp.~2862--2896, 2021.

\bibitem{Sellentin:2016psv}
E.~Sellentin and A.~F. Heavens, ``{Quantifying lost information due to
  covariance matrix estimation in parameter inference},'' {\em Mon. Not. Roy.
  Astron. Soc.}, vol.~464, no.~4, pp.~4658--4665, 2017.

\end{thebibliography}


\end{document}